\documentclass{aa}  

\usepackage{graphicx}
\usepackage{txfonts}
\usepackage{chemfig}
\usepackage[version=4]{mhchem}
\begin{document}

   \title{Monte Carlo simulation of UV-driven synthesis of complex organic molecules on icy grain surfaces}


   \author{Y. Ochiai
          \inst{1}
          \and
          S. Ida
          \inst{1}
          \and
          D. Shoji\inst{2}
          }

    \institute{Earth-Life Science Institute, Tokyo Institute of Technology, Meguro-ku, Tokyo 152-8550, Japan \\
              \email{ochiai@elsi.jp}
         \and
         Institute of Space and Astronautical Science, Japan Aerospace Exploration Agency, Chuo-ku, Sagamihara, Kanagawa 252-5210, Japan
           }

   \date{DRAFT:  \today}

 
  \abstract
   {Complex organic molecules (COMs) have been widely observed in molecular clouds and protostellar environments.
   One of the formation mechanisms of COMs is radical reactions on the icy grain surface driven by UV irradiation. While many experiments have reported that various COMs (including biomolecules) can be synthesized under such ice conditions, the majority of the reaction processes are unclear. Complementary numerical simulations are necessary to unveil the  synthetic process behind the formation of COMs.}
   {In this study, we develop a chemical reaction simulation using a Monte Carlo method. To explore the complex reaction network of COM synthesis, the model was designed to eliminate the need to prepare reaction pathways in advance and to keep computational costs low. This allows for broad parameter surveys and a global investigation of COM synthesis reactions.
   With this simulation, we investigate the chemical reactions occurring on icy dust surfaces during and after UV irradiation, assuming a protoplanetary disk environment.
   We aim to reveal the types of organic molecules produced in a disk and the formation mechanisms of COMs, in particular, amino acids and sugars.}
   {The  Monte Carlo method we developed here produces reaction sequences by selecting a reaction from all candidate reactions at each calculation step, based on Arrhenius-type weighting.
   For our purpose, we significantly accelerated the calculation by adopting an approximate estimation of activation energy without expensive quantum chemical calculations.}
   {The results show that photodissociation and subsequent radical-radical reactions cause random rearrangement of the covalent bonds in the initial molecules composed of methanol, formaldehyde, ammonia, and water. Consequently, highly complex molecules such as amino acids and sugars were produced in a wide range of the initial conditions. 
   We found that the final abundances of amino acids and sugars have extremely similar dependence on the atomic ratios of the initial molecules, which peak at $\rm C/H \sim 0.1$-0.3 and $\rm O/H \sim 0.3$-0.5, although the amino acids abundance is usually more than ten times higher than that of sugars. To understand this dependence, a semi-analytical formula was derived.
   Additionally, parameter surveys of temperature, photon energy, and other  factors have suggested that the decomposition reactions of amino acids and sugars undergo a rapid transition within the threshold of a given parameter.
   }
   {}


   \maketitle
%

\section{Introduction}\label{sec:intro}
Recent developments in observational techniques have allowed diverse organic molecules to be observed in molecular clouds and protostellar environments.
\citep[e.g.,][]{Herbst2009,Yang2022,Sewilo2022-ey,Rocha2024}.
Complex organic molecules (COMs) are generally defined as molecules containing six or more atoms, which includes at least one carbon atom, such as methanol, glycolaldehyde, or urea.
The presence of COMs in space indicates that rich chemistry should take place prior to and over the course of star formation.
The various organic molecules discovered in samples of comets \citep{McKay2021}, meteorites \citep{Sephton2002,Pizzarello2010}, and asteroid Ryugu \citep{Naraoka2023,Yabuta2023} may be related to the COMs originating from molecular clouds and/or protoplanetary disks.
Therefore, COM formation is a key stage for chemical evolution in space.
They are thought to be formed mainly in ice mantles of dust grains \citep[e.g.,][and references therein]{oberg2016,PP7_ceccarrelli}.
The ice phase is generally unfavorable for chemical reactions due to low temperature.
However, in environments where icy dust is exposed to cosmic rays, ultraviolet light, and/or X-ray irradiation, these high-energy particles break the chemical bonds of the surface molecules and create radicals, which are highly reactive \citep{Arumainayagam2019}.
They initiate radical reactions and generate various molecules, even under cryogenic temperatures.
Many laboratory experiments have been conducted using different energy sources, such as UV lights, X-rays, and high-energy electrons, with various initial materials, confirming the synthesis of COMs in ice \citep[e.g.,][]{Sugahara2019, Bulak2021,Munoz_Caro2019,Habershon_2015}.

\citet{ciesla2012} suggested that some of dust particles in a protoplanetary disk are temporarily exposed to UV irradiation due to the disk gas turbulence transporting them up to the upper layer.
If icy grains, which are in the outer region of the disk, are lifted, COM synthesis could be caused on the ice dust surface through radical reactions driven by UV irradiation.
While observational constraints on the synthesis of COMs within disks are currently insufficient, \citet{Yamato2024-hs}  recently reported the detection of COMs in a disk by observations with Atacama Large Millimeter/submillimeter Array (ALMA), suggesting the potential for COM synthesis in protoplanetary disks.
COMs that are synthesized in a disk could also serve as a source of organic molecules in meteorites, potentially contributing to Earth's prebiochemistry thereafter.

The fundamental and crucial questions for the COMs' synthesis in the icy mantles of dust grains are what types of organic molecules are synthesized and how they are produced in the reaction networks.
To date, a wide variety of COMs have been identified in many laboratory experiments mimicking the ice-surface reactions in astrochemical environments \citep[e.g.,][and references therein]{oberg2016,PP7_ceccarrelli}.
While some studies have shown that highly complex molecules, such as amino acids, sugars, and nucleobases, can be synthesized by UV irradiation to simple organic ice, they are generally detected in ex situ analyses \citep[e.g.,][]{Munoz2002,Meinert2016,Oba2019}.
In other words, those molecules are found by analysing the refractory residues obtained after the ice samples are heated to room temperature.
Such procedures could lead to additional chemical reactions in the samples, making it difficult to determine when the products were formed and what their original structures were.

While some highly effective in situ analysis techniques using time-of-flight mass spectrometers have been developed in recent decades \citep[e.g.,][]{Gudipati2012,Henderson2015,Turner2020,Turner2021}, which are successful in identifying many COMs, including toluene ($\rm C_7H_8$), acetamide ($\rm CH_3CONH_2$), glycolaldehyde ($\rm HCOCH_2OH$), ethynamine ($\rm HCCNH_2$), and other species, 
it is still challenging to specifically determine complexly structured COMs such as amino acids, sugars, and nucleobases, as well as to comprehensively analyze the products.
Furthermore, radical reactions, which are important for COMs synthesis, cause complex reaction networks and diverse products.
Thus, elucidating all of the products of COMs synthesis by only experiments is a demanding undertaking. In addition, reaction processes are basically inaccessible in laboratory experiments.
Although infrared (IR) spectrometers enable the monitoring of the presence and the concentration change of relatively simple molecules and radicals during the reaction \citep[e.g.,][]{Munoz2003,Zhu2021, Martin2020} without destructive processing, it is difficult to interpret overall complex reaction networks from only those results.
Therefore, as a complementary approach, a numerical simulation would be necessary to unveil the  synthetic process behind COM formation, alongside laboratory experiments.

In the present paper, we tackle the investigation of the global chemical reactions leading to COM synthesis and the prediction of the dependence of yields of COMs (especially amino acids and sugars on initial molecules) using a newly developed forward Monte Carlo simulation code.
We aim to reveal the types of COMs produced on the ice-dust surface in a protoplanetary disk irradiated by UV from the host star, and their formation mechanisms.
Importantly, this study does not focus on searching for synthetic pathways of a specific molecule backwardly (i.e., from a target molecule to reactants) because such an approach does not allow capture of the global reaction processes.

Thus, we did not adopt a reaction network model, which assumes the reaction pathways in advance and solves the rate equations \citep[e.g.,][]{chang_2016, garrod_2019, jin_2020} in this study.
This model is a powerful tool for investigating the formation process of relatively small COMs (and also large ones whose synthesis pathways are obtainable) and the parameter dependence with low calculation costs; however,  the reaction networks involving relatively large COMs, such as amino acids and sugars, are mostly unknown and would be overly complicated.
We attempt to explore the synthesis of COMs by including these molecules in a forward method (i.e., from given reactants to products), which theoretically predicts reaction pathways. 

For the purpose of simulating chemical reactions a priori without assuming the pathways, quantum chemical calculations were also used \citep[e.g.,][]{Goldman2010,saitta_2014,Zamirri2019,Inostroza2019,Sameera2022}.
However, they are computationally too expensive to handle a large number of atoms and long timescales, which are important for studying the synthesis process of COMs.
Conducting extensive parameter surveys using them is not practical either.
Therefore, it is necessary to use a method that can theoretically generate reaction networks and has low calculation costs at the same time. 
With this motivation, we developed a Monte Carlo simulation for chemical reactions for the purposes of this study. 

As a prototype of our model, we used the simulation method proposed by \citet{takehara}.
These authors investigated the sugar synthesis on icy grain surface in a protoplanetary disk using their original Monte Carlo simulation and proposed a new synthesis process of sugars: formation of macromolecules by UV irradiation followed by the break-down after the irradiation.
The methodology in this study follows the basic idea of the prototype scheme, but also adds new effects and modifications to reproduce more realistic reaction networks. Details of individual formalism are explained in the method section (Sect.~\ref{sec:method}).

Here, we briefly summarize the basic structure of the prototype scheme and our new modifications.
As the site of organic molecule synthesis, icy grains in a turbulent protoplanetary disk are considered. The grains are irradiated by UV light from the host star when they are vertically stirred up and shielded otherwise. Additionally, warm ($T\sim 50$-100 K) regions of a disk are the objective environment, where molecular diffusion on the grain surface is not severely restricted.
The mathematical descriptions of molecules and chemical reactions are based on the graph-theoretical matrix-type ``DU model” proposed in \citet{dugundji}.
 For the Monte Carlo calculation, in each calculation step, one reaction is selected from all candidates derived from an ensemble of molecules (EM), using the weighted probabilities. Repeating this selection, we proceed through a chemical reaction sequence. A statistically significant number of reaction sequences are calculated, starting from the same initial EM and using different random numbers.
We select reactions based on activation energy ($E_{\rm a}$) as follows: 1) the enthalpy change ($\Delta H$) in a reaction is derived by the bond energy change of EM before and after the reaction; 2) adopting the Bell-Evans-Polanyi principle, $E_{\rm a}$ is evaluated from $\Delta H$. 
For the purposes of our study, we added the following modifications: 
3) the selection probability of each candidate reaction is weighted by the Arrhenius-type equation with $E_{\rm a}$; 4) using the Eyring equation, which relates $E_{\rm a}$ with the reaction timescale, we prevent the selection of very slow reactions. 
Finally, the radical reactions and CO, which play  important roles in organic chemical reactions, were explicitly formulated in this study. We note that in \citet{takehara} the former was too simplified and the latter was not included.
    
This method reduces calculation costs by utilizing bold approximations in the procedures in the selection steps described above, instead of carrying out quantum chemical calculations.
This allows for simulations of complex reaction networks and broad parameter surveys that are not usually achievable with conventional approaches.
Although the yields of individual products and reaction pathways cannot be predicted with high accuracy, the approximations are allowable for our purpose 
in this study. 
We conduct calculations using various initial molecules and different parameters, aiming for a comprehensive understanding of the COMs' synthesis processes.

To mimic the ice-surface reaction in a protoplanetary disk, the simulations are conducted with simple organic molecules such as H$_2$O, NH$_3$, HCN, CH$_3$OH, and CH$_2$O as the initial molecules, below temperatures of 100 K.
The complex reaction networks initiated by photodissociation reactions is simulated.
This study also focuses on the synthesis of amino acids and sugars, which are biomolecules and have been found in meteorites \cite[e.g.,][for a review]{Ehrenfreund2001,Koga2017,furukawa_2019}, and Ryugu samples \citep{Parker2023}, investigating the relationship between their final abundances and atomic ratios of the initial molecules. Furthermore, the dependence on parameters such as temperature and photon energy is examined.

In Sect.~\ref{sec:method}, we explain our Monte Carlo simulation scheme. Since our model is original and conceptually new, a detailed description 
is given.
In Sect.~\ref{sec:result}, we apply our model for COM synthesis occurring during and after UV irradiation. 
In our simulation framework, the reaction process is dominated by photodissociation and radical-radical reactions, which cause the randomization of covalent bonds within the initial molecules.
Based on this mechanism, we show that atomic ratios of C/H and O/H of the initial molecular set are key parameters for the final abundance of amino acids and sugars.  
In Sect.~\ref{sec:discussion}, we discuss comparison with experimental results, observations, and implications for COM synthesis in protoplanetary disks.  
We summarize our results in Sect.~\ref{sec:conclusion}.

\section{Method}
\label{sec:method}

\subsection{The site of synthesis: Surface of icy grains in the ``warm'' regions of the disk} \label{sec:icy_grain_surface}

As explained in Sect~\ref{sec:intro}, this study focuses on the chemical reactions that would occur on the surface of icy dust grains in protoplanetary disks.
In general, protoplanetary disks are too dense for UV light to penetrate the inner region.
However, 
grains are occasionally stirred by the disk gas turbulence to the upper layer and exposed to UV radiation from the host star \citep{ciesla2012}. 
As grains grow and become less coupled with the turbulent gas, their scale height decreases; in other words, the diffusion to the upper layers becomes inefficient.
Although some grains may experience multiple transitions between the exposed and shielded areas, many grains once exposed to UV light would eventually settle into the disk without a further shift.

Therefore, in this simulation, two phases are prepared to mimic such a situation:
1) UV phase, corresponding to the period when grains are exposed to UV irradiation in the upper layer; and 2) post-UV phase, corresponding to the period after grains sink into the disk and are shielded from UV light. The molecules are exceedingly activated during the UV phase and undergo relaxation in the post-UV phase. 
Once molecules experience strong activation, they can end up with a completely different molecular distribution from the initial state, even after the energy supply is terminated.

In interstellar molecular clouds, $T$ is so low ($\sim$10--30 K) that the molecular diffusion would highly depend on the type of molecules, the composition of ice, the temperature and so on \citep[e.g.][]{jin_2020}.
The present study focuses on relatively warm ice (50-100 K), corresponding to the disk regions beyond Jupiter's orbit up to Neptune's orbit, where molecules may diffuse easily, and the dependence of diffusion on species could be small.
\citet{tachibana} showed that amorphous ices composed of $\rm H_2O, CH_3OH,$ and $\rm NH_3$ and of pure $\rm{H_2O}$ exhibit liquid-like behavior within temperature range of 65-150 K and 50-140 K, respectively.
This phenomenon would enhance the random collisions among ice molecules in a warm disk region further.

Therefore, in this study, we assume that molecules diffuse and interact independently of the species and neglect differences in the diffusion energies of individual molecules.
This assumption allows us to list the possible reactions from a given collection of molecules with simple conditions (see~Sect.~\ref{sec:list_candidates}).
Under this assumption, it is demonstrated that the reaction networks of the chemical system computed in this study do not significantly change in the temperature range of 40-130 K, as discussed in Sect.~\ref{sec:d_tem}.

\subsection{DU model; graph-theoretic matrix model for chemical reactions}\label{sec:DU}

Our simulation uses the DU model \citep{dugundji}, which is a graph-theoretical matrix-type model, for the mathematical description of molecules and chemical reactions. 
The DU model is one of ``logic-oriented'' chemical reaction models, which were actively discussed in 1970's and 80's, mainly for the backward search for synthesis pathways of target molecules for industrial purposes, while they were not frequently considered after 1980's.
Since astrochemistry often takes place in extreme environments (i.e., cryogenic temperatures, low pressure, and intense UV flux), which significantly differ from laboratory conditions, the logic-oriented model is useful for our purpose.

The fundamental unit of the DU model is an ensemble of molecules (EM) consisting of $n$ atoms; $A = {A _1, A_2, ..., A_n}$.
The model focuses on the chemical reactions within an EM.
Molecules are characterized by the constituent atoms and the connection between them.
In other words, atoms correspond to vertices, and chemical bonds correspond to edges in graph theory.
With the DU model, chemical reactions are reduced to algebraic mathematics, which is directly incorporated into Monte Carlo simulations detailed in this paper.
Here, we describe only the matrix representation of molecules and chemical reactions. 
The method for incorporating it into Monte Carlo simulations is explained in Sect.~\ref{sec:MonteCarlo}.

\begin{figure*}
  \centering
  \includegraphics[width=14cm]{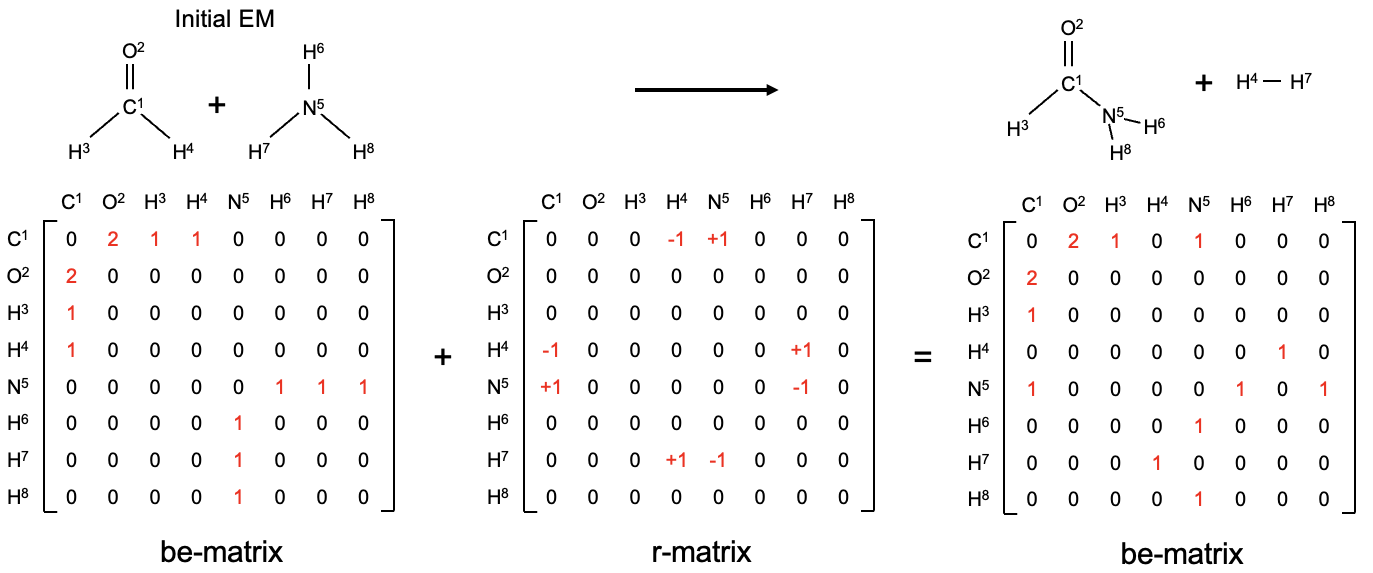}
  \caption{Example of a chemical reaction between formaldehyde and ammonia molecules represented by a be-matrix and an r-matrix of the DU model.}
  \label{fig:matrix}
\end{figure*}

An EM is expressed by a matrix called ``be-matrix'' (bond and electron) that is an $n \times n$ symmetric matrix.
The $i$-th row (and $i$-th column) is assigned to $A_i$, the $i$-th atom constituting the EM. 
The entry $b_{ij}$ ($i\ne j$) represents the bond order between $A_i$ and $A_j$, in other words, the number of valence electrons pairs between $A_i$ and $A_j$. 
The entry $b_{ii}$, a diagonal entry, represents the number of unshared valence electrons of $A_i$ in the original DU model.
However, it is always given zero in our calculations, as those electrons making a lone pair are not involved in the reactions, except for the new treatment for CO molecules developed in this paper.
We also stress that since ion reactions are not considered in this study, atoms are always neutral, so, the number of valence electrons does not change (regarding radicals, see Sect~\ref{sec:radicals}).
To summarize, the be-matrix satisfies the following conditions in this study:
\begin{align} 
1. & \:\: b_{ij}\ge 0, \label{eq:be-matrix1} \\ 
2. & \sum_{1\le j\le n} b_{ij} = \mbox{the valency of }A_i. 
\label{eq:be-matrix2} 
\end{align}

Figure~\ref{fig:matrix} shows the example of a be-matrix represented by the DU model. 
Any isomeric EMs that are composed of the same collection of atoms, but of different molecules can be represented by changing the bond orders and/or their locations in the matrix.
Since a chemical reaction is an exchange of chemical bonds between molecules, 
a chemical reaction is expressed by the addition of the reaction matrix, ``r-matrix,'' which represents the transfer of bonds. 
While the $i$-th row (and $i$-th column) of reaction matrix corresponds to the atom $A_i$ as well as in the be-matrix, its entry $r_{ij}$ ($i\ne j$) represents how many covalent bonds are made (+) or broken (-) through the chemical reactions.
Thus, adding the r-matrix to the starting state be-matrix gives the post-reaction be-matrix.
In this study, following \citet{takehara}, the hydrogen atoms are specially grouped into a single column ($m$-th column) by summing the number of hydrogen atoms connecting $A_i$ and placing it in the entry $b_{im}$, in order to reduce the calculation costs, although this is not shown in Fig.~\ref{fig:matrix}.

\subsection{Monte Carlo simulation with the DU model}
\label{sec:MonteCarlo}
    This section describes the methodology of the prototype Monte Carlo simulation scheme for chemical reactions proposed in \citet{takehara} and the updates made in this study.

    \subsubsection{Listing all candidate reactions}
    \label{sec:list_candidates}
    
    In this model, one simulation step of chemical reactions is limited to the exchange of two chemical bonds, which is the smallest chemical change.
    In other words, two bonds, whether within a single molecule or across two different molecules, are broken and new bonds are formed, resulting in the production of new molecules. 
    Double bonds and triple bonds, such as ones in oxygen molecules and nitrogen molecules are treated as isolated two and three bonds in this calculation.
    Therefore, in the cleavage of these bonds, double bonds are converted to single bonds, and triple bonds to double bonds, respectively.
    This restriction to a chemical reaction allows us to derive all possible reactions within a given EM by considering all pairs of two bonds.
    We stress that this restriction is not an idea of the original DU model, but was introduced to fit with the forward Monte Carlo simulation in \citet{takehara}.
Based on the conditions for the be-matrix and the constrain on chemical reactions above, the conditions for the r-matrix are given as follows:
1) The matrix is composed of two entries with -1 (bond cleavage), two entries with +1 (bond formation), their diagonally symmetric entries and all other entries with 0 (except for reactions involving CO); 2) the entries with -1 cannot be assigned to the position of $b_{ij}=0$. They cannot be located in the same row and column either; 3) the entries with +1 are assigned to the row and column which correspond to the atoms whose bonds are broken. (In other words, $ \sum_{1\le j\le n} r_{ij} = 0$.)
By generating r-matrices under these three conditions, all potential reactions that can occur within an EM are therefore listed.

In the calculations conducted in this study, an EM typically yields 100-1000 candidate reactions. 
We interate to select one reaction from them to proceed through a reaction sequence that comprises 3000 steps (see Sect.~\ref{sec:weighting}).
From the same initial EM, $10^5-10^6$ distinct reaction sequences are calculated  (Sect.~\ref{sec:abundance}).

The reaction networks computed using this method can be too huge for reaction network models to prepare in advance. Currently, only Monte Carlo simulation is available to address this problem.

   
\subsubsection{Weighting probabilities of reactions}
\label{sec:weighting}

After candidate reactions are listed in the procedure above, one reaction is randomly chosen based on the weighted probabilities that reflect reaction rates.
The equation of the weight is given as follows:

\begin{equation}
    W = \exp{\left(-\frac{E_{\rm a}}{\mathrm{R}T}\right)} \,,
    \label{eq:W-weight}
\end{equation}
where $\mathrm{R}$ is the gas constant, $T$ is temperature, and $E_{\rm a}$ is the activation energy of the reaction.
This equation is based on the Arrhenius equation that provides reaction rate constants.
The denominator of the exponent in Eq.~(\ref{eq:W-weight}) is written as 
\begin{align}
\mathrm{R}T \simeq 0.83 \left(\frac{T}{100\, \rm K}\right) \:\: \rm kJ/mol.
\end{align}
Under temperature of $T = 100\,\rm K$: a fiducial temperature in this study, a difference in activation energy ($E_{\rm a}$) of only 10 kJ/mol produces a significant change in a weight ($W$) by a factor of $1.7 \times 10^5$. 
As the reaction rates are exceedingly sensitive to the dependence on $E_{\rm a}/T$, the pre-exponential factor is ignored for simplicity.
The evaluation of $E_{\rm a}$ is discussed in Sect.~\ref{sec:BEP}.

Based on the probabilities weighted by Eq.~(\ref{eq:W-weight}), a generated random number specifies one reaction from the candidates.
While the reaction with the maximum $W$ is not necessarily selected, reactions with large $W$ are selected with higher probability. 
After one reaction is chosen, the new EM after the reaction is designated as the next starting molecules. 
Then, the reaction step, consisting of listing reaction candidates, weighting the probabilities, and selecting one reaction, is repeated to advance through subsequent reactions.

\subsubsection{Derivation of abundances of individual molecules}
\label{sec:abundance}

In this study, we follow the chemical reaction sequence that consists of 2800 steps in the UV phase, followed by 200 steps in the post-UV phase (see Sects.~\ref{sec:icy_grain_surface} and \ref{sec:radicals}).
Using the same initial molecules, the reaction sequence is computed $N_{\mathrm{run}} = 10^5-10^6$ times with a different set of random numbers.
A large number of reaction sequences samples provide the molecular distribution in each reaction step and its evolution.



Following \citet{takehara}, the ``abundance'' of molecules and structures, such as functional groups, in the $i$-th step is defined as follows:
\begin{equation}
\label{eq:abundance}
    A_i=n_{\mathrm{target}(i)}/N_{\mathrm{run}},
\end{equation}
where 
$n_{\mathrm{target}(i)}$ is the number of a target molecule or structure at the $i$-th step in the total $N_{\mathrm{run}}$ runs of calculations.
Thus, $A_i$ represents the average number of targets that exist in the molecular set of our simulations (same as an EM of the DU model) in the $i$-th step, and depends on the size of the prepared molecular set (the total number of atoms in the set).
Since our current simulation does not have spatial information, the abundance cannot be directly translated into values such as concentration or column density of molecules in ice.
However, the abundance ratios between molecules are comparable to the ratios of these values.

As will be shown later (Sect.~\ref{sec:postUVphase}), we constrain the timescale of reactions occurring in the simulation. If the activation energy for all candidate reactions is too high to occur within that timescale, the chemical reaction is finally stalled. Therefore, after the reaction cessation, the quenched abundance of molecules is referred to as the final abundance ($A_{\rm final}$).


\subsection{Selecting reactions based on activation energy} \label{sec:selection}
\subsubsection{Evaluation of $E_{\rm a}$ from the enthalpy change}\label{sec:BEP}

\begin{figure}
  \centering
  \includegraphics[width=6cm]{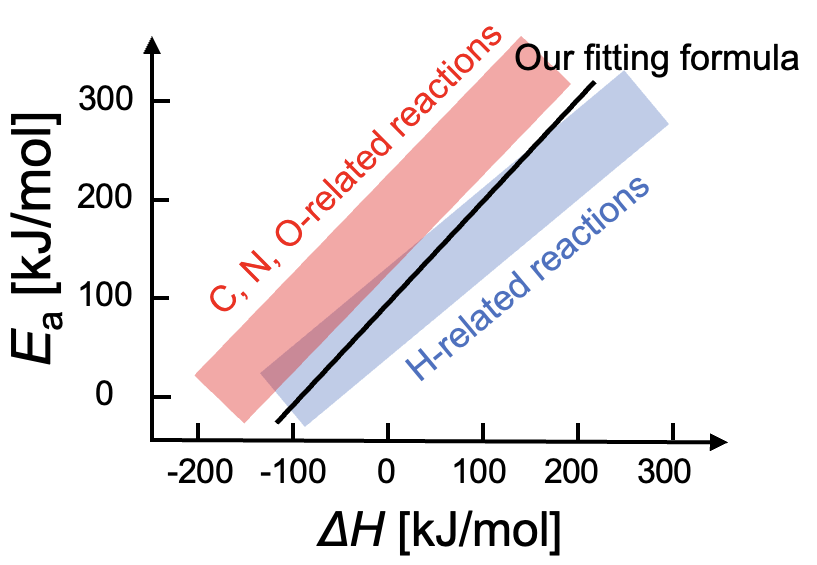}
  \caption{
  Schematic diagram of the linear relationship between the enthalpy change and activation energy predicted by the Bell-Evans-Polanyi principle.
  While there are variations in each reaction, many of the reaction types considered in this study are shown to be broadly classified into two pairs of $\alpha$-$\beta$ \citep[e.g.,][]{michaelides_2003, Wang_2011_evans,sutton_2012}, as illustrated with the red and blue regions.
}
  \label{fig:BEPrelation}
\end{figure}

To investigate global reaction networks and general trends depending on initial and environmental conditions, rather than detailed analysis of individual reactions, this method employs the Bell-Evans–Polanyi principle to evaluate the activation energy.
This principle describes the relationship between the activation energy, $E_{\rm a}$, and the enthalpy change, $\Delta H$, before and after a reaction:
    \begin{equation}
    \label{eq:evans}
        E_a=\alpha \, \Delta H+\beta, 
    \end{equation}
where $\alpha$ and $\beta$ are empirical parameters given by $\alpha \sim 0.7$-1 and $\beta \sim 80$-$180 \,\rm kJ/mol$ from density functional theory (DFT) calculations \citep[e.g.,][]{michaelides_2003, Wang_2011_evans,sutton_2012}. 
These calculations also have shown that many of the reaction types considered in this study are broadly classified into two pairs of $\alpha$-$\beta$ as illustrated in Fig.~\ref{fig:BEPrelation}. 
However, in this work, we always assume one specific pair of $\alpha$ and $\beta$ in each calculation for simplicity. As the representative values, $\alpha = 1$ and $\beta = 100 \,\rm kJ/mol$ are used.
 Although the Bell-Evans–Polanyi principle is an approximate formula and contains uncertainties in the values of $\alpha$ and $\beta$, these variations do not significantly affect the reaction process simulated in this study.
However, the progress of the decomposition reaction of amino acids and sugars could depend on the value of $\beta$. (see detailed discussion in Sect.~\ref{sec:d_ab}).


\subsubsection{Evaluation of the enthalphy change from bond energy} \label{sec:dH}

\begin{table}[hbtp]
  \caption{Bond energy used in this study from \citet{sanderson}. }
  \label{table:bond}
  \centering
  \begin{tabular}{lr|lr}
    \hline
      & Bond energy  &  & Bond energy  \\
    \hline \hline
    C $-$ C  & 347 kJ/mol  & N $-$ N  & 161 kJ/mol\\
    C $=$ C  & 611 kJ/mol   & N $=$ N  & 456 kJ/mol\\
    C $\equiv$ C & 837 kJ/mol & N $\equiv$ N  & 946 kJ/mol\\
    C $-$ N  &  305 kJ/mol  & N $-$ O  & 230 kJ/mol \\
    C $=$ N  & 615 kJ/mol  & N $=$ O  & 598 kJ/mol\\
    C $\equiv$ N  & 891 kJ/mol  & N $-$ H  & 389 kJ/mol\\
    C $-$ O  & 358 kJ/mol  & O $-$ O  & 146 kJ/mol\\
    C $=$ O  & 745 kJ/mol  & O $=$ O  & 498 kJ/mol\\
    C $\equiv$ O &1072 kJ/mol &  O $-$ H  & 464 kJ/mol\\
    C $-$ H  & 414 kJ/mol  & H $-$ H  & 426 kJ/mol\\
    \hline
  \end{tabular}
\end{table}
The enthalpy change, $\Delta H$, is calculated by the change in the sum of bond energy of the molecules before and after the reaction. 
For the calculation of $\Delta H$, fixed bond energy values for each bond, as shown in Table~\ref{table:bond}, are used.
Although bond energies can vary due to molecular species and, under icy conditions, to interactions with surrounding molecules, they are neglected in this study.
As will be shown in Sec.~\ref{sec:UVphase}, the reactions simulated in this study are dominated by UV-induced photodissociation and radical-radical reactions, both of which would not be quite sensitive to such a bond energy change. 
Furthermore, in a reaction system containing a huge variety of molecular species focused on in this study, a species-dependent deviation of bond energies may be averaged out.

    If the equation provides negative $E_{\rm a}$, in other words,
    \begin{equation}
    \label{eq:zeroEa}
        \Delta H<-\frac{\beta}{\alpha},
    \end{equation}
    we set $E_{\rm a}=0$ to avoid unphysically large $W$ (Eq.~(\ref{eq:evans}))
    \footnote{
    Since the reactions with $E_{\rm a}<0$ are barrierless, the differences in $\Delta H$ should not cause differences in the reaction rate. 
    In this case, direct substitution of $E_{\rm a}$ calculated by Eq.~(\ref{eq:evans}) into Eq.~(\ref{eq:W-weight}) is not relevant.    
    \citet{takehara} used the enthalpy change instead of the activation energy in Eq.~(\ref{eq:evans}) without a treatment for the enthalpy change corresponding negative $E_{\rm a}$. 
    This treatment with Eq.~(\ref{eq:evans2}) is newly added in our study.
    }.
    Thus, our prescription is
    \footnote{This modified form is comparable to the more detailed version of the Bell-Evans–Polanyi principle where $E_{\rm a}\rightarrow 0$ and $\alpha \rightarrow 0$ for smaller vales of $\Delta H, $ namely, at the the exothermic limit
    \citep[e.g.,][]{Santen2010}.}
    \begin{align}
    \label{eq:evans2}
        E_a & = \max\left(0, \; \alpha \, \Delta H+\beta\right).
    \end{align}

\subsubsection{Timescale limitation based on activation energy} \label{sec:t_Ea}

The time length of a reaction step in this calculation does not represent a specific size, but changes depending on the frequency of chemical changes occurring in the simulating real system.
The lower the frequency of chemical changes, the longer the time length corresponding to each reaction step.
Thus, even in one series of a reaction sequence, it can vary.

The reaction selection method described above is based on only the relative probabilities among the candidate reactions, not considering absolute magnitudes of reaction rates. One problem is that even if the activation energies of all the candidates are too high to occur in the simulated system, one of them will always be selected. To avoid this, in the prototype model \citep{takehara}, the reaction candidate of no chemical change with $\Delta H=0$ was prepared in each step.

In this paper, we adopt a more logical prescription. Since grains experience vertical movement by turbulent diffusion, we restrict the reactions to occur on timescales longer (with high $E_{\rm a}$) than the typical circulation timescale of grains between the upper, UV-exposed layer and the lower, shielded layer.
       
    The Eyring equation, based on the transition state theory, gives the reaction rate constant as follows:
    \begin{equation}
    \label{eyring1}
        k=\frac{\kappa k_B T}{h} \exp{\left(-\frac{\Delta G^\ddag}{RT}\right)}
         =\frac{\kappa k_{\rm B} T}{h} \exp{\left(\frac{\Delta S^\ddag}{R}\right)}\exp{\left(-\frac{\Delta H^\ddag}{RT}\right)},
    \end{equation}
    where $k$ is the rate constant, $\Delta G^\ddag$ is the Gibbs energy of activation, $\kappa$ is the transmission coefficient, $k_{\rm B}$ is the Boltzmann constant, R is the gas constant, $T$ is the temperature, and $h$ is the Planck constant.
    Assuming that the $\kappa \simeq 1$ and the entropy change $\Delta S^\ddag$ is negligible in ice reactions,
    $k$ is rewritten as:
    \begin{align}
    \label{eyring2}
        k \simeq & 2.084 \times 10^{12} \left(\frac{T}{\mathrm{100 \, K}}\right) \nonumber\\ & \times \exp{\left\{-\left(\frac{\Delta H^\ddag}{0.8314\,\mathrm{kJ/mol}}\right)\left(\frac{T}{\mathrm{100\, K}}\right)^{-1}\right\}} \;\rm s^{-1}
    \end{align}
    Replacing $\Delta H^\ddag$ in the transition state with the activation energy $E_a$, the half-life timescale of reactants can be written as:
    \begin{align}
    \label{logk}
        \log_{10}\left(\frac{t_{1/2}}{\mathrm{year}}\right) 
        \simeq &
        \log_{10} \left( \frac{\ln 2}{k/\mathrm{s}^{-1}} \right)
        \simeq-0.159-\log_{10}(k/\mathrm{s^{-1}}) \nonumber\\
         \simeq & -19.97-\log_{10}\left(\frac{T}{\mathrm{100\,K}}\right) \nonumber\\
        & +20.90
        \left(\frac{E_{\mathrm{a}}}{40\,\mathrm{kJ/mol}}\right)\left(\frac{T}{\mathrm{100\,K}}\right)^{-1},
    \end{align}
    This equation is solved regarding the activation energy as:
    \begin{equation}
    \label{eq:Ecrit}
    \begin{split}
        E_{\mathrm{a,crit}}&\simeq 40\,\mathrm{kJ/mol} \times \left(\frac{T}{\mathrm{100\,K}}\right) \times \\
        &\qquad \left[ 1.10 + 0.048 \log_{10} \left\{ \left( \frac{t_{1/2}}{\mathrm{10^3year}} \right) \left( \frac{T}{\mathrm{100\,K}} \right) \right\} \right].
    \end{split}
    \end{equation}
From this equation, we obtain the critical activation energy $E_{\mathrm{a,crit}}$, which corresponds to the upper limit of the reaction timescale $t_{1/2}$.
In this paper, $t_{1/2} = 1000 \, \rm years$ is given as the grain vertical circulation timescale \citep[e.g.,][]{Klarmann2018}.
In the Monte Carlo scheme, a reaction candidate involving no chemical change with $E_{\mathrm{a}}=E_{\mathrm{a,crit}}$ is set at each step. Reactions with $E_{\rm a} > E_{\mathrm{a,crit}}$, which occur on longer timescales than the assumed environmental conditions, are less likely to be selected.
As a result, molecules in the calculations finally tends to stay at a local minimum state.

\subsection{Newly incorporated ``bonds''}

The prototype model by \citet{takehara} adopted the simple prescription that all the valence electrons of each atom are used to form covalent bonds.
In this study, we newly incorporate radicals and CO molecules into the DU model and the Monte Carlo scheme.

\subsubsection{Introduction of radicals} \label{sec:radicals}

As mentioned in Sect~\ref{sec:icy_grain_surface}, we set the UV irradiation phase before the following post-UV phase (shielded phase). The UV phase, which is very important for the organic synthesis on the icy grain surfaces, is characterized by the occurrence of radical reactions by UV photons.

One significant improvement from the prototype model is the explicit introduction of radical reactions.
Radical species were not represented in the original DU model \citep{dugundji} either.
Since \citet{takehara} considered only molecules forming covalent bonds, they converted the photon energy to temperature to represent radical reactions induced by photons. 
They assumed that reactions at high temperatures correspond to the UV phase reactions and all molecules exist by repeatedly undergoing photodissociation and immediate recombination.
While random reactions independent of $E_{\rm a}$ caused at extremely high temperature could partly reflect the extreme destruction by photons and barrireless radical-radical reactions, we need an explicit prescription to simulate UV irradiation and radical reactions, considering the actual environmental temperatures.

In this study, we develop a methodology to represent radical reactions that align with the DU model, integrating it into our Monte Carlo simulation. Specifically, we introduce a hypothetical element ``X'' to express radical species. A bond with an X, X-R (R = C, N, O, or H), are not a covalent bond, but represents the presence of an unpaired electron on the atom. For example, a hydroxyl radical, $\rm \cdot OH$, is expressed as $\mbox{X-OH}$.
An X bond is treated as a single bond, same as a H bond, but with zero bond energy. Next, to express a photodissociation reaction, a hypothetical molecule ``X$_2$'' is introduced. As an example, photodissociation of a $\rm H_2O$ molecule,
\begin{align}
{\rm H_2 O} + h\nu \rightarrow \rm H \cdot + \cdot OH,
\end{align}
is represented as
\begin{align}
\mbox{H-O-H} + \mbox{X-X} \rightarrow \mbox{H-X} + \mbox{X-O-H}.
\label{eq:XOH}
\end{align}
For a photodissociation reaction, ``hypothetical negative bond energy of X$_2$'' is employed to represent an energetic UV photon cleaving covalent bonds. 
This is based on the principle that the lower the bond energy, the greater the reactivity of the bond.
Accordingly, to represent the input of a photon energy of $E_{\rm UV}$, the negative bond energy of $-1000 \,(E_{\rm UV}/10 \,{\rm eV}) \:\: \rm kJ/mol$ is assigned to X$_2$. 
We always set one X$_2$ in the molecular set at every step in the UV phase to mimic continuous UV irradiation. In the post UV phase, the supply of X$_2$ is stopped.
Finally, for radical-radical reactions such as a reverse reaction of Eq.~(\ref{eq:XOH}), the produced X$_2$ is removed from the molecular set.
The bond energy of the produced X$_2$ is set to zero, ignoring the heat generated by the radical recombination. 
This is based on the assumption that the heat is immediately radiated at longer wavelengths.

 The above prescription for a photodissociation reaction allows our simulation to describe the transition from high-energy UV irradiation (which breaks covalent bonds) to low-energy irradiation (which does not). More details on this are given in see Sect.~\ref{sec:d_photonE}.
The supply and consumption of X$_2$ molecules by photodisociation reactions and radical-radical reactions eventually make the system go toward an equilibrium between the production and extinction of radicals in the UV phase.
On the other hand, radicals decrease as a reaction step proceeds in the post UV phase because of the cessation of X$_2$ supply as mentioned in the prescription 4.

The enthalphy change of a photodissociation reaction is written as follows:
\begin{align}
\Delta H_{\mathrm{pd}} = BE_{\mathrm{broken}} -1000 \,\left(\frac{E_{\rm UV}}{10 \,{\rm eV}} \right)\:\: \rm kJ/mol,
\end{align}
where $BE_{\mathrm{broken}}$ represents the bond energy of the broken bond.
Since the bond energies used in this study are $< 470 \, \rm kJ/mol$ (Table~\ref{table:bond}), $\Delta H_{\mathrm{pd}}$ with $E_{\rm UV}=10\rm \,eV$ (fiducial photon energy) gives a negative value to any type of bond.

The probabilities of the $\mathrm{X}_2$-involved reactions being selected are also weighted based on Eq.~(\ref{eq:W-weight}).
However, the activation energy $E_{\mathrm{a}}$ of a radical-radical reaction is always set to zero without using the Bell-Evans-Polanyi equation (Eq.~(\ref{eq:evans})).
While some radical-radical reactions have been found to have activation barriers with a few $\rm \, kJ/mol$ presumably by the influence of surrounding water molecules \citep{Enrique-Romero2022-mv}, we do not consider them because of the little effects on results in this study.
For photodissociation reactions, as the molecule is directly activated by a photon breaking the bond, we use $\Delta H$ for weighting instead of activation energy, as follows:
\begin{equation}
        W = \exp{\left(-\frac{\Delta H_{\mathrm{pd}}}{\mathrm{R}T}\right)} \,.
        \label{eq:X-weight}
\end{equation}
If $\Delta H_{\mathrm{pd}}$ is negative, we set $\Delta H_{\mathrm{pd}}=0$, same as in the evaluation with $E_\mathrm{a}$.

In reality, however, the dissociation rate is not determined solely by the magnitude of the bond energy.
Some molecules, such as $\mathrm{N_2}$, $\mathrm{H_2}$, and $\mathrm{CO}$, hardly dissociate even if the photon energy is larger than the bond energy.
For CO molecules, we adopt an exception: we set a barrierless radical reaction as a dissociation process instead of photodissociation (see Sect.~\ref{sec:CO}).
    

\subsubsection{Introduction of carbon monoxide}\label{sec:CO}

\begin{figure}
  \centering
  \includegraphics[width=5cm]{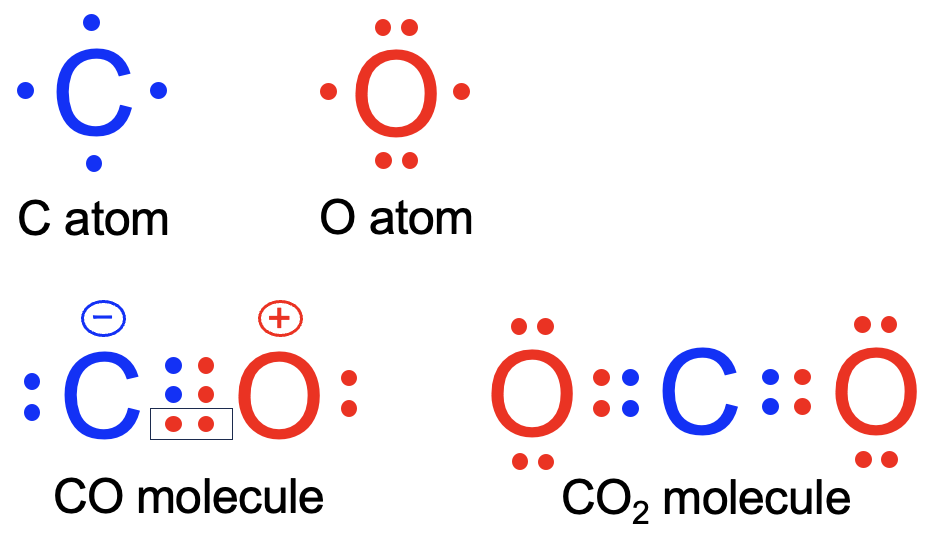}
  \caption{Lewis structures of a carbon atom, an oxygen atom, a carbon monoxide molecules and a carbon dioxide molecule.
  In the CO molecule, the lone pair of the oxygen atom shared with the carbon atom and the electrical charge imbalance are illustrated.
  This structure can be expressed by a be-matrix (Sect.~\ref{sec:CO}).
}
  \label{fig:CO}
\end{figure}

While carbon monoxide is quite ubiquitous in space and is an important molecule for astrochemistry, 
it was not included in the previous model by \citet{takehara} because 
of its bond property. 
In this study, we newly introduce the prescription for CO.

Unlike the bond in which two atoms contribute one electron each, a CO triple bond is formed by the sharing of a lone pair from the oxygen atom (Fig.~\ref{fig:CO}).
Thus, when the CO triple bond is cleaved and becomes a double bond, instead of one unpaired electron being distributed to each atom, two unpaired electrons are distributed in the carbon atom.
This cleavage allows the carbon atom to bond with two more atoms. The reaction of CO therefore can be written as follows:

\begin{equation}\label{eq:COreaction}
    \chemfig{C~O + R-R'}
    \longrightarrow
    \chemfig{O=C(-[:60]R)-[:-60]R'}
\end{equation}

To express CO using the DU model, the diagonal entry of C and O in the be-matrix are set to be +1 and -1, respectively. The entry of the bond order between C and O atoms is set to be 3 to conserve the condition 2 for a be-matrix (Eq.~(\ref{eq:be-matrix2})).
This prescription was not considered in the original DU model.

The reaction between two CO molecules is prohibited in this study because the product, ethylene dione (O=C=C=O) is known to be extremely unstable \citep{Schroder1998}.
The CO-forming reaction, which is the reverse of Eq.~(\ref{eq:COreaction}), occurs when the C-R and C-R' bonds within the O=CRR' molecule are broken.

The photodissociation of CO is also prohibited in this study since the triple bonds of CO cannot be cleaved by ultraviolet light with an energy of 10 eV \citep{Heays2017}.
While the reaction between CO and a radical, on the other hand, is incorporated  according to Eq.~(\ref{eq:COreaction}), it is not selected in the calculations due to the activation barrier.
However, the reactions of CO molecules in ice have been investigated over decades, and it has been found that CO can undergo several reactions even at cryogenic temperatures, such as 10 K.
Whereas the reaction pathways have not been fully identified, the following reaction has been suggested to have no barriers \citep{Oba2010}:
\begin{equation}\label{eq:COpath}
    \ce{CO + \cdot OH -> CO2 + H \cdot}.
\end{equation}
Therefore, in this study, the activation energy of this reaction is arbitrarily given zero, instead of the calculated values using Eq.~(\ref{eq:evans}).
We note that this is a tentative setting for reaction pathways of CO and will be updated.
\citet{Molpeceres2023} found that this reaction has an activation barrier when taking place in amorphous solid water (ASW) and in CO ice due to stabilization of the intermediate, HOCO.
However, since the reaction process of CO is not yet fully understood at the moment, and we confirmed that changing the CO reaction pathways has little effect on the results of this study, Eq.~(\ref{eq:COpath}) is used in this study.

\section{Results}\label{sec:result}

To understand an overall picture of the reactions during and after UV irradiation, we first focus on the types of reactions that take place and the associated changes in bonding and molecular size in each of the UV and post-UV phases.
Then the synthesis of amino acids and sugars as two characteristic complex organic molecules are investigated.
Finally, dependence on the parameters used in this calculation will be shown.

\subsection{Initial molecules and simulation settings}\label{sec:fiducial}

Hereafter, unless otherwise noted, 2 methanol molecules (CH$_3$OH), 5 formaldehyde molecules (CH$_2$O), 9 ammonia molecules (NH$_3$), and 22 water molecules (H$_2$O) are used as the initial molecules.
These molecules are commonly found in interstellar space \citep[e.g.,][and references therein]{Gibb2004-ei}.
The simulation parameters adopted in a fiducial set are:  
the disk temperature $T=100\,\rm{K}$, the UV photon energy of 10 eV (equivalent to X$_2$ energy of $-10^3 \rm \, kJ/mol)$, and the numerical factors of the Bell-Evans–Polanyi principle, $\alpha=1$ and $\beta=100 \, \rm{kJ/mol}$ (Eq.~(\ref{eq:evans})).
The dependence on the initial molecules, $T$, the photon energy, $\alpha$ and $\beta$ are discussed in Sects.~\ref{sec:semiAS}, \ref{sec:d_tem}, \ref{sec:d_photonE}, and \ref{sec:d_ab}.

The UV phase is applied for reaction step 1 to 2800, and the post-UV phase is applied for reaction step 2801 to 3000.
The total steps of the UV phase are large enough for the molecules to reach an equilibrium state where the formation and extinction of all bond-types between C, N, O, H and X balance, as shown below.
We repeat $10^6$ runs from the same initial molecules with different random number sequences in the fiducial case.

\subsection{Reactions in the UV phase}\label{sec:UVphase}

\begin{figure}
  \centering
  \includegraphics[width=8.5cm]{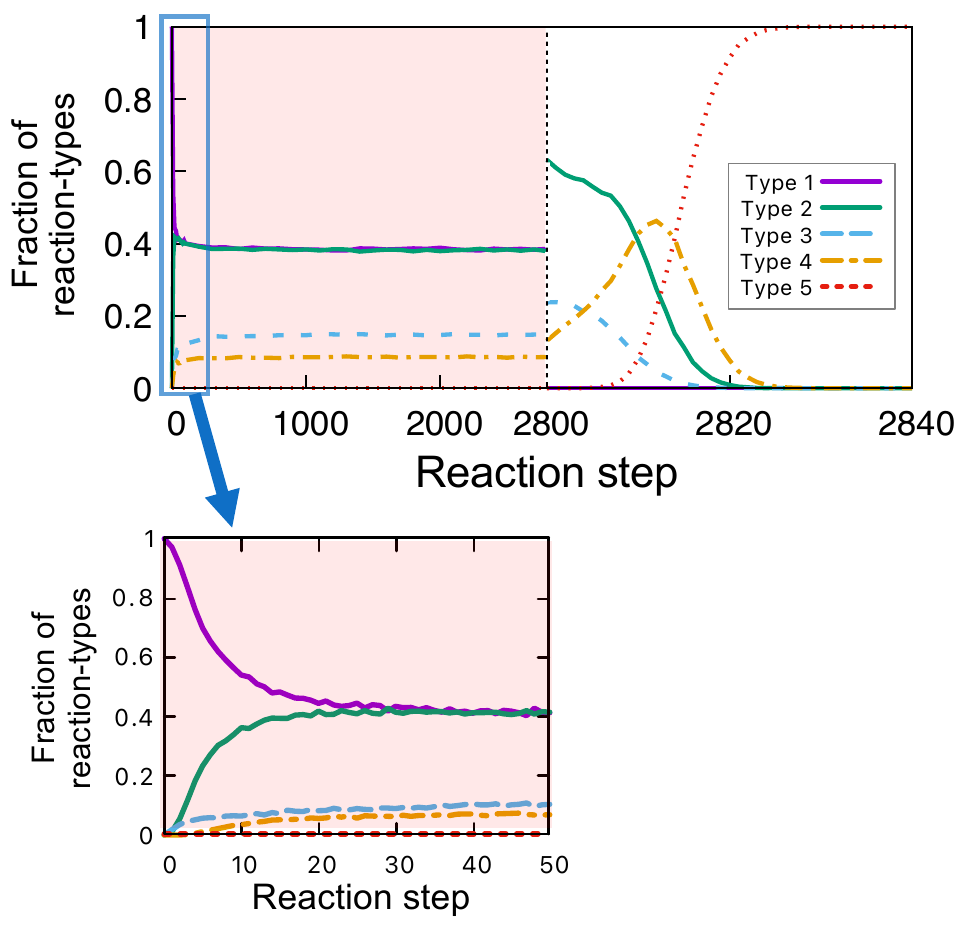}
  \caption{Fraction change in the five reaction types defined in Sect.~\ref{sec:UVphase}.
  The parameters and initial molecules used are shown in Sect.~\ref{sec:fiducial}.
  Reaction type 1, 2, 3, 4, and 5 are represented by the magenta solid, green solid, light-blue dashed, orange dash-dotted, and red doted lines, respectively. 
  The red shaded region corresponds to the UV phase. The bottom figure is a stretched view of up to step 50. The horizontal scales are different between the UV phase and the post-UV phase for visual convenience. 
}
  \label{fig:reactiontype}
\end{figure}

\begin{figure}
  \centering
  \includegraphics[width=7.5cm]{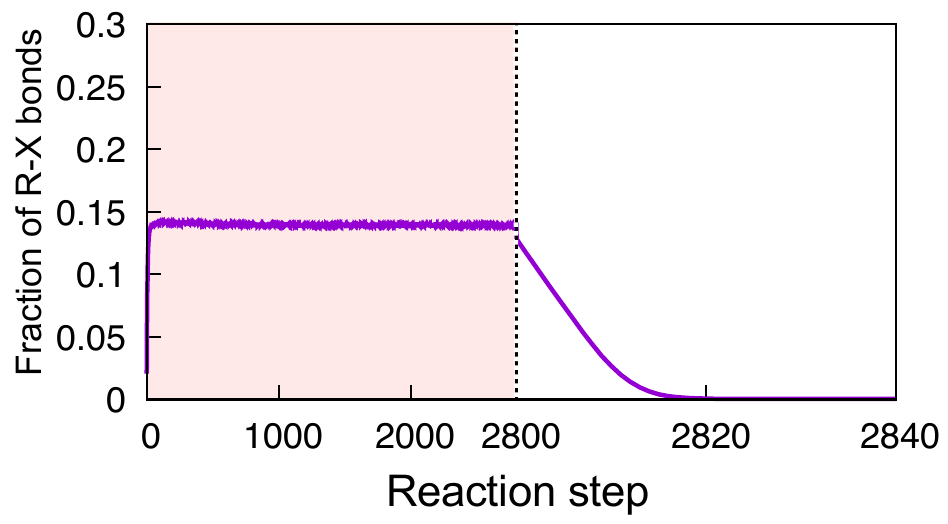}
  \caption{Change in fraction of R-X bonds (R=C, N, O, or H atoms) to total bonds.
  The parameters and initial molecules used are described in Sect.~\ref{sec:fiducial}.
}
  \label{fig:radical}
\end{figure}

\begin{figure}
  \centering
  \includegraphics[width=7.5cm]{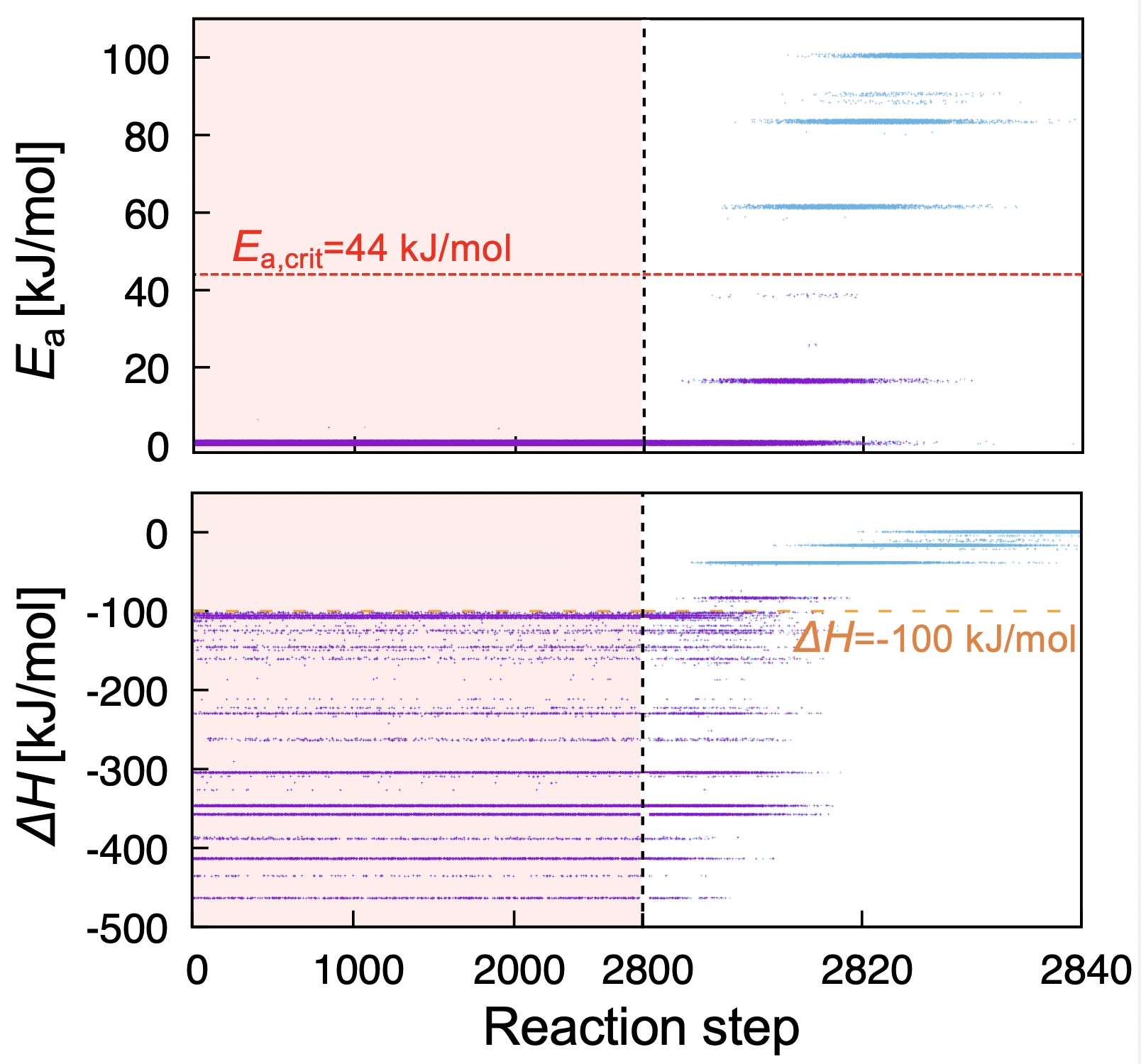}
  \caption{
 Activation energy, $E_{\rm a}$, (upper panel) and the enthalpy change, $\Delta H,$ ( lower panel) of  type 2, 3, and 4 reactions selected in each step.
 The red dashed line shows the critical activation energy ($E_{\rm a, crit}=44\,\rm kJ/mol$) under the fiducial timescale and temperature. 
  The orange dashed line shows the threshold ($\Delta H=-100\, \rm km/mol$) which gives $E_{\rm a, crit}=0$ under $\alpha =1.0$ and $\beta =100\, \rm kJ/mol$ (see Sect.~\ref{sec:dH}).
  For visual convenience, we arbitrarily selected the number of plotting data and randomly shifted the points vertically and horizontally with the small amplitude. 
  The magenta dots are results with $E_{\rm a,crit} = 44 \rm \, kJ/monl$ and the light blue dots are results from the same initial conditions but without limiting $E_{\rm a,crit}$.
  They show that the activation energy of the selected reaction is limited by the $E_{\rm a,crit}$ setting. We note that in most steps the light blue dots are hidden by the magenta dots.
  }
  \label{fig:reactionEa}
\end{figure}

The COM synthesis simulated in this study is initiated by radical reactions.
To analyze this process, we classify chemical reactions into five types as shown in Table~\ref{table:reactiontype}
\footnote{
In ice conditions, reactions 3 and 4, which involve the cleavage of chemical bonds, could be inhibited because the energy of the reactants can dissipate during the formation of intermediates \citep[e.g.,][]{Martinez-Bachs2023-sz,Pantaleone2021-tn}. 
However, since this study assumes an environment in which continuous UV irradiation is occurring, even if the reaction stops at an intermediate state, bond dissociation would proceed subsequently due to direct or indirect influences of UV photons. 
At present, due to a lack of general understanding of stabilization processes in ice reactions, we do not consider the energy transport between molecules.
}.

{\renewcommand\arraystretch{1.2}
\begin{table}[hbtp]
\centering
  \caption{Five reaction types considered in this study. A, B, C, and D represent carbon, nitrogen, oxygen, or hydrogen atoms, and X is the hypothetical element used to represent radical reactions (see Sect.~\ref{sec:radicals}).}
  \label{table:reactiontype}
  \begin{tabular}{l|l}
  \hline
  Type 1 & Photodissociation\\
   &  (A-B + X-X $\rightarrow$ A-X + B-X) \\
  \hline
  Type 2 & Radical-radical reaction\\
   &  (A-X + B-X $\rightarrow$ A-B + X-X) \\
  \hline
  Type 3 & Radical-nonradical reaction\\
   &  (A-X + B-C $\rightarrow$ A-B + C-X)\\
  \hline
  Type 4 & Nonradical-nonradical reaction\\
   & (A-B + C-D $\rightarrow$ A-C + B-D)\\
  \hline
  Type 5 & No change\\
   &  (reaction pathway with $E_{\rm{a}}= E_{\rm{a,crit}}$) \\
  \hline
  \end{tabular}
\end{table}
}

Figure~\ref{fig:reactiontype} shows the fraction change in the reaction types during the UV and post-UV phases as a function of reaction steps.
At the beginning of the UV phase, photodissociation (type 1) dominates the reactions, converting initial molecules into radicals.
As radicals increase, the fraction of type 2 and 3, in which radicals act as reactants, rises.
Eventually, radical formation (type 1) and extinction (type 2) are balanced after a few dozens of reaction steps.
The equilibrium is also established in the fraction of the radical bonds to the total bonds (Fig.~\ref{fig:radical}).

For the type 1 reactions with a photon energy of 10 eV, the $\Delta H$ is always negative (Eq.~(\ref{eq:X-weight})) \footnote{Even if the total enthalpy change of the photodissociation reaction is negative according to the calculation using the hypothetical negative bond energy of X$_2$, the dissociated molecule is exceedingly destabilized (the actual $\Delta H$ is $>0$).} because the equivalent energy of 1000 kJ/mol is large enough to cut any types of a covalent bond of C, N, O and H (Table~\ref{table:bond}).
This means that those reactions are calculated as the same weights ($W = 1$) as reactions with $E_{\rm a}=0$.
Type 2 (radical-radical) reactions are the most powerful opposing reactions to type 1 reactions and are always barrierless ($E_{\rm a} = 0$) with $W = 1$ (Eq.~(\ref{eq:evans2})).

Type 4 reactions also occur at about 10\% probability, even though no radicals are involved.
They are mainly the reactions of C-H bond, an abundant structure, with other unstable bonds such as O-O, N-O, and N-N.
Since these reactions are highly exothermic (for example, $\Delta H$ of ``C-H + O-O $\rightarrow$ C-O + O-H'' is -262 kJ/mol using the bond energies from Table~\ref{table:bond}), they do not have activation barriers ($E_{\rm a}=0$) in most ranges of $\alpha$ and $\beta$ based on the Bell-Evans-Polanyi principle (Eq.~(\ref{eq:evans2})).
Those unstable bonds cannot be thermally formed under this temperature, but are formed by radical-radical reactions.
Therefore, while radicals are not directly involved in type 4, they are driven by UV irradiation as well as the radical reactions (type 1, 2, and 3).

Figure~\ref{fig:reactionEa} shows the $E_{\rm a}$ and $\Delta H$ of the selected reactions (type 2, 3 and 4) in typical runs in the fiducial set.
It shows that almost all the selected reactions in the UV phase are barrierless ($E_{\rm a}=0$). In other words, only the reactions with $\Delta H \le -\beta/\alpha = -100 \, \rm kJ/mol$ (Eq.~(\ref{eq:evans2})) are selected from the reaction candidates,
This is due to the continuous photodissociation (type 1) shown in Fig.~\ref{fig:reactiontype}, providing an abundant supply of energetically-unstable radicals.
Such a situation is realized in the upper UV-exposed layers of the disk.



Since type 1 and 2 reactions, which are dominant in the UV phase, occur independently of the bond-type, they cause random rearrangements of the covalent bonds between C, N, O, and H atoms, providing various bond-types as shown in Fig.~\ref{fig:bondchange}.
Type 3 and 4, on the other hand, produce stable bonds from relatively unstable bonds, while the reverse reaction does not proceed due to the high activation barrier.
Hence, the final bond distribution is regulated by the abundance ratios between C, N, O, and H atoms and the stability of the bonds,
rather than the forms of initial molecular species.

\begin{figure*}
  \centering
  \includegraphics[width=14cm]{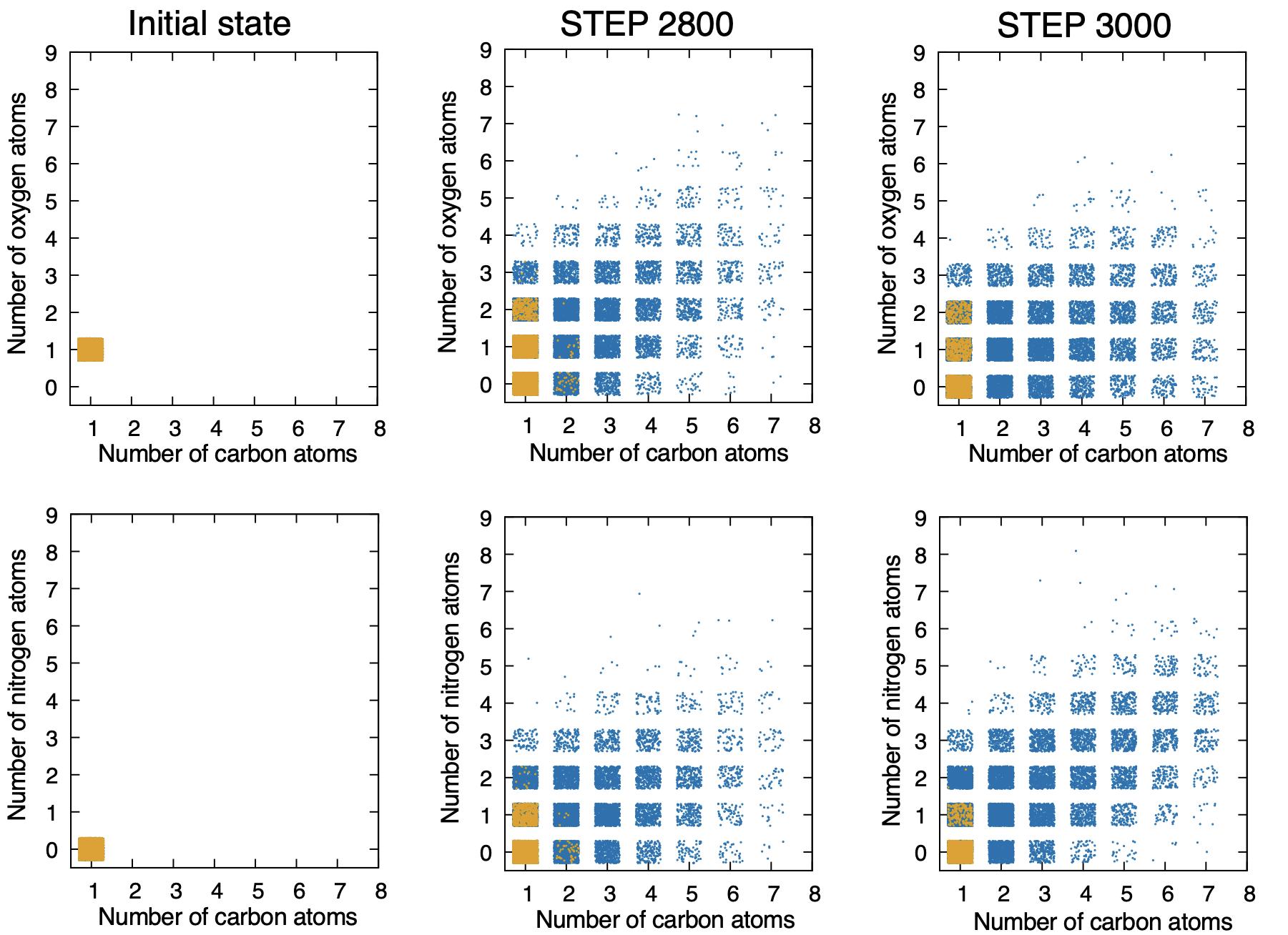}
  \caption{
  Distributions of the numbers of carbon atoms and oxygen atoms (the upper panels) and carbon atoms and nitrogen atoms (the lower panels) included in an individual molecule,
  at the initial state, the end of the UV phase, and the end of the post UV phase from the left to the right, respectively.
  The results are obtained with 5000 runs.
  The blue dots represent the molecules with six or more atoms (COMs), and the orange dots represent others.
  We note that relatively small COMs such as methanol are hidden behind the orange dots (including at the initial state.)
  For visual convenience, the points are randomly shifted in the range of $\pm 0.3$ from the center. 
  }
  \label{fig:molecularsize}
\end{figure*}

Through the random rearrangement,the bonds such as C-C, C-N, and N-N are newly formed (Fig.~\ref{fig:bondchange}), resulting in the growth of complex molecules that consist of six or more atoms as shown in Fig.~\ref{fig:molecularsize}.
It should be stressed that the biggest size of molecules cannot be predicted only from this result because it depends on the number of atoms included in the initial molecules in our simulations, that allow all atoms to interact freely.

\subsection{Reactions in the post-UV phase}\label{sec:postUVphase}

\begin{figure*}
  \centering
  \includegraphics[width=18cm]{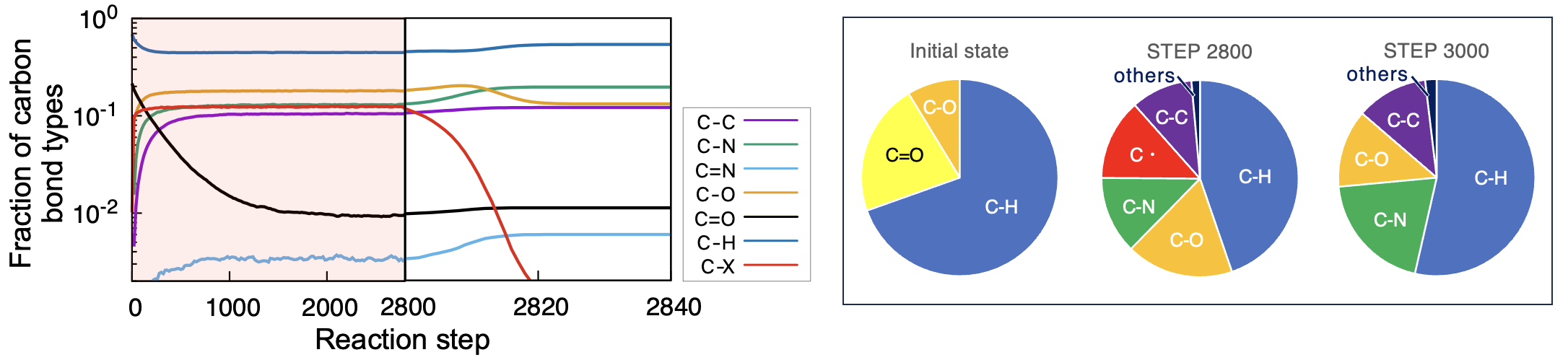}
  \caption{Fraction change in carbon bond types as a function of reaction steps with a logarithmic scale (the left panel) and its pie chart (the right panel)  
  at the initial state, the end of the UV phase, and the end of the post UV phase from the left to the right, respectively. In the left panel, the bond-types with less than 0.1 \% throughout the reaction steps are excluded.
  }
  \label{fig:bondchange}
\end{figure*}

Once the UV phase ends, the photodissociation stops, and radical-radical reactions become dominant (Fig.~\ref{fig:reactiontype}). Most of the radicals produced in the UV phase are consumed in the first 20 steps (Fig.~\ref{fig:radical}) and randomly form new covalent bonds.
However, if unstable bonds like N-N, N-O, and O-O are formed, they are barrierlessly degraded through mainly the reactions with C-H bonds, making more stable bonds with hydrogen or carbon atoms. 
This process selectively replaces C-H bonds with C-N and C-O bonds. While the decrease in C-H bond is canceled by other reactions, the slight increases in C-N and C-O fractions are shown in Fig.~\ref{fig:bondchange}.
While a hydrogen atom can only bond with one other atom, nitrogen and oxygen atoms can form bonds with multiple atoms.
Therefore, when an oxygen or nitrogen atom is attached to a single carbon bond, the number of atoms attached to it is always greater than when a hydrogen atom is attached.
The increase of C-N and C-O bonds thus enhances the complexity of carbon molecules.


As the barrierless reactions decrease, reactions with activation barriers ($E_{\rm a}=10$-40 kJ/mol) start to take place (Fig.~\ref{fig:reactionEa}). 
The typical reactions with activation barriers observed in the fiducial calculation are 
\begin{align}
\mbox{C-O + H-H} \rightarrow \mbox{C-H + O-H} & \:\:\:\: (\Delta H = -84 \, \rm kJ/mol),
\label{eq:COHH}
\end{align}
and, in a smaller portion of
\begin{align}
\mbox{C-N + H-H} \rightarrow \mbox{C-H + N-H} & \:\:\:\: (\Delta H=-62 \,\rm kJ/mol), \label{eq:CNHH} \\
\mbox{C-O + C-H} \rightarrow \mbox{C-C + O-H} & \:\:\:\: (\Delta H=-39 \, \rm kJ/mol).
\label{eq:COCH} \end{align}
Reactions ~(\ref{eq:COHH}) and ~(\ref{eq:CNHH}) strip O and N atoms from C atoms, providing C-H bonds (Fig.~\ref{fig:bondchange}).
This process competes with the radical-radical reactions that form various carbon bonds, as well as the reactions above that increase C-N and C-O bonds, in terms of influencing the complexity of carbon molecules.

In sufficiently H-rich cases, most of the molecules end up bonding to only hydrogen atoms such as CH$_4$, NH$_3$, and H$_2$O.
The complex molecules generated during the UV phase are just decomposed by hydrogenation after UV irradiation stops.
However, in most initial conditions that we examined in Sect.~\ref{sec:semiAS},
including this fiducial case, molecular complexity is maintained or enhanced in the post-UV phase. 

In the post-UV phase, the molecules in the system are gradually stabilized.
As the molecules become more stable, $E_{\rm a}$ of the possible next-step reactions tend to become higher, meanig that the reactions slow down in reality.
As explained in Sect.~\ref{sec:t_Ea}, this study imposes a limitation on reaction timescales of $10^3 \rm \, years$.
This timescale is converted to the critical activation energy $E_{\rm a,crit} = 44 \, \rm kJ/mol$ (Eq.~(\ref{eq:Ecrit})), which inhibits the reactions with higher activation energies (the light-blue dots in Fig.~\ref{fig:reactionEa}) from being selected.
Therefore, once the low-barrier reactions with $E_{\rm a}<E_{\rm a,crit}$ are exhausted, further reactions hardly occur.
Figures~\ref{fig:reactiontype} and \ref{fig:reactionEa} show that the fraction of a type 5 reaction ($E_{\rm a}=E_{\rm a,crit}$) increases with the upward shift of the selected $E_{\rm a}$, indicating that the reactions are finally quenched.

As explained in Sect.~\ref{sec:d_time}, the magnitude of $E_{\rm a,crit}$ heavily depends on temperature, and higher temperatures result in more degradation of C-N and C-O bonds, leading to the less diverse molecules.
Hence, 
the complex molecules are conserved without decomposition due to the low temperature.

\subsection{Definition of ``initial'' molecules} \label{sec:meaningofIM}

In this calculation, all the initial molecules experience dissociation due to the sufficient number of
reaction steps in the UV phase and are randomly reconnected.
They thus lose information about their initial molecular forms, and the final molecular distribution is almost entirely determined by the atomic ratio of C, N, O, and H atoms, which is conserved throughout the reaction sequence.

However, in the real systems, the atomic ratio is not necessarily conserved 
due to composition changes (adsorption and/or desorption of molecules) on the ice surface.
The composition changes during the post-UV phase may have little effect on the reaction networks since barrierless radical-radical reactions are dominant and are completed within the first few dozens of steps.
Hence, the reaction processes of the post-UV phase can be solely determined by the molecular distribution just before UV irradiation stops.
This allows our simulations to predict the final products even if the absorption and desorption processes during the UV phase are unknown.
At the same time, the initial molecules here 
should be considered to give the atomic composition of the final state of the UV phase.

\subsection{Synthesis of amino acids and sugars}

\subsubsection{Formation processes}

\begin{figure}
  \centering
  \includegraphics[width= 8.5cm]{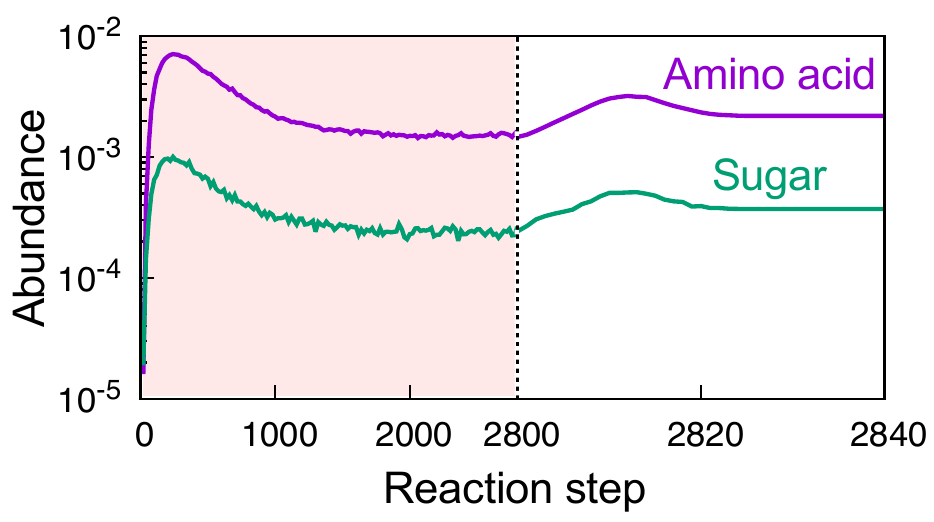}
  \caption{
  Evolution of the abundances of amino acids (magenta curve) and sugars (green curve) as a function of reaction steps.
  The parameters and initial molecules used are the fiducial set shown in Sect.~\ref{sec:fiducial}.
  }
  \label{fig:p3aminosugar}
\end{figure}

\begin{figure}
  \centering
  \includegraphics[width=8.5cm]{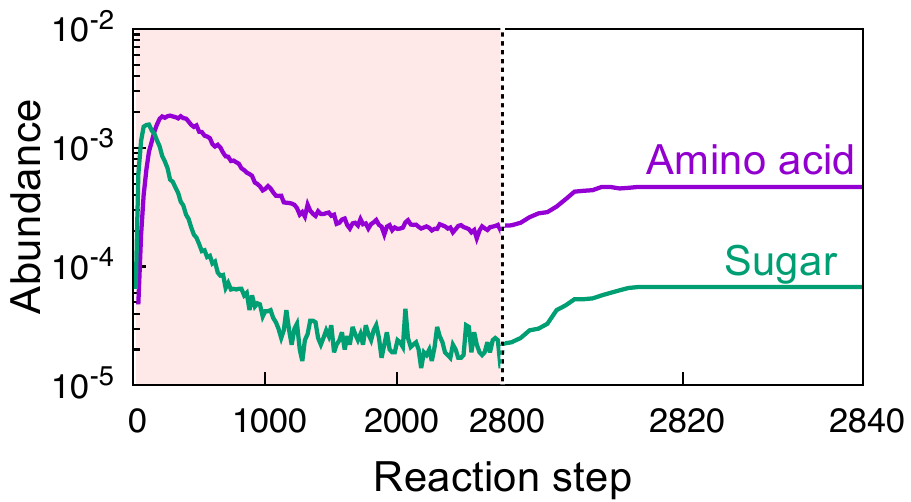}
  \caption{
  Same plot as Fig.~\ref{fig:p3aminosugar} except for the initial molecules: 1 methane molecule ($\rm CH_4$), 18 formaldehyde molecules ($\rm CH_2O$), 5 hydrogen cyanide molecules ($\rm HCN$), and 1 ethylene molecule ($\rm C_2H_4$).
  }
  \label{fig:p44aminosugar}
\end{figure}

\begin{figure}
  \centering
  \includegraphics[width=8.5cm]{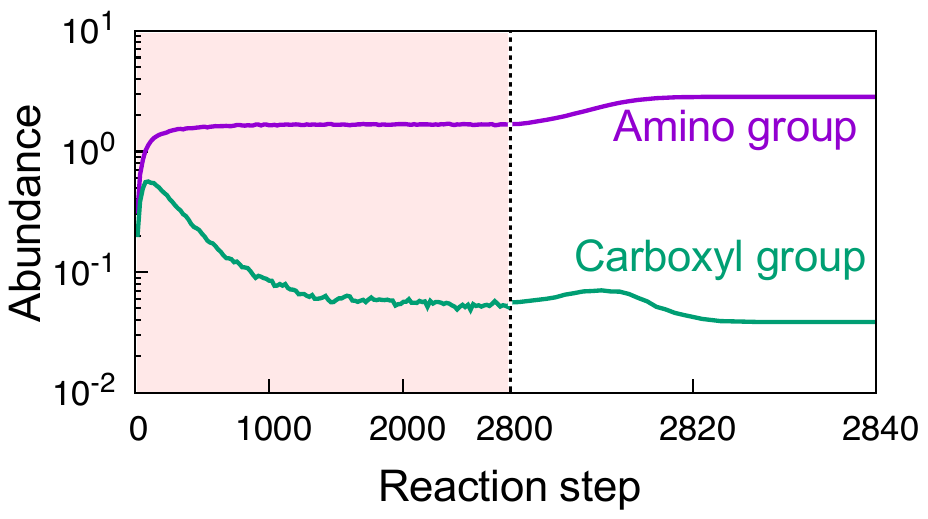}
  \caption{Change in the abundances of amino group (magenta curve) and carboxyl group (green curve) in the fiducial calculation. Only amino groups bonded to carbon atoms ($\rm C$-$\rm NH_2$) are counted in this plot. 
  }
  \label{fig:group}
\end{figure}

Figure~\ref{fig:p3aminosugar} shows the change in the abundances of amino acids and sugars in the fiducial set.
In this study, amino acids and sugars refer to $\alpha$-amino acids and aldoses, respectively.
An aldose has three or more carbon atoms, and its ring structure is also included.
The 
abundances of amino acids and sugars quickly elevate at first and reach equilibrium after 2000 steps. 
In the post-UV phase, they increase slightly and decomposition continues; however, it stops immediately as the reaction is quenched around step 2830 (see Sect.~\ref{sec:postUVphase}).


While the abundance of sugars are about one order of magnitude less than that of amino acids throughout the reaction steps, they exhibit extremely similar behavior.
The similarity is found in many other sets of initial molecules, although some show the different evolution in the first few hundred steps due to the significant dependence of individual molecular species production on the structure of the initial molecules (Fig.~\ref{fig:p44aminosugar}).

For amino acids and sugars, the Strecker reaction and the formose reaction have conventionally been proposed as synthetic routes, respectively \citep[e.g.][] {Singh2022,Danger2011, Fresneau2015, Meinert2016}.
However, in our simulation, these build-up-type pathways are not the major routes, as molecules are constantly broken apart and randomly reassembled  by radical-radical reactions. 
Rather than following a specific pathway, in the simulated conditions, the molecules stochastically form the structure of amino acids or sugars according to the randomization of the structures.

To analyze the synthesis of amino acids based on this view point, the abundances of amino groups and carboxyl groups, the main building blocks of amino acids, are plotted as a function of reaction steps in Fig.~\ref{fig:group}. 
It shows that the abundance of amino groups is more than one order of magnitude higher than that of carboxyl groups after step 1000.
This indicates that the carboxyl group production is a bottleneck for the amino acid synthesis.
We note that as the reaction proceeds, the structures are highly randomized between molecules, and it does not happen that a particular molecule exclusively dominate a certain structure.

On the other hand, despite the absence of carboxyl groups in the sugar structure, the evolution of the sugar abundance shows a pattern very similar to that of amino acids.
The common structure between a carboxyl group and a sugar is a C=O bond, 
which suggests that they both are regulated by the production of C=O bonds.
It should be noted that most of the sugar molecules produced in our simulations are chain-form sugars, not cyclic sugars that do not have C=O bonds.

Based on this idea, we attempt to reduce synthetic processes of amino acids and sugars into the production of the individual bonds constituting them.
Even if different initial molecules are used, the formation of constitutive bonds through random bond rearrangement and destruction by photodissociation should compete and form these molecules.
At the same time, because the C=O bond is the hardest to be formed among all bonds in amino acids and sugars across a broad range of initial molecular sets, including the fiducial case (Fig.~\ref{fig:bondchange}), their final abundances tend to be controlled by the amount of C=O bonds.

We show in Sect.~\ref{sec:semi_analy} that the abundance distribution of amino acids and sugars in broad range of initial atomic ratios is reasonably and consistently explained by the hypothesis that the C=O bond is a key factor.

\subsubsection{Dependence on initial C/H and O/H ratios}\label{sec:semiAS}

\begin{figure}
  \centering
  \includegraphics[width=8.5cm]{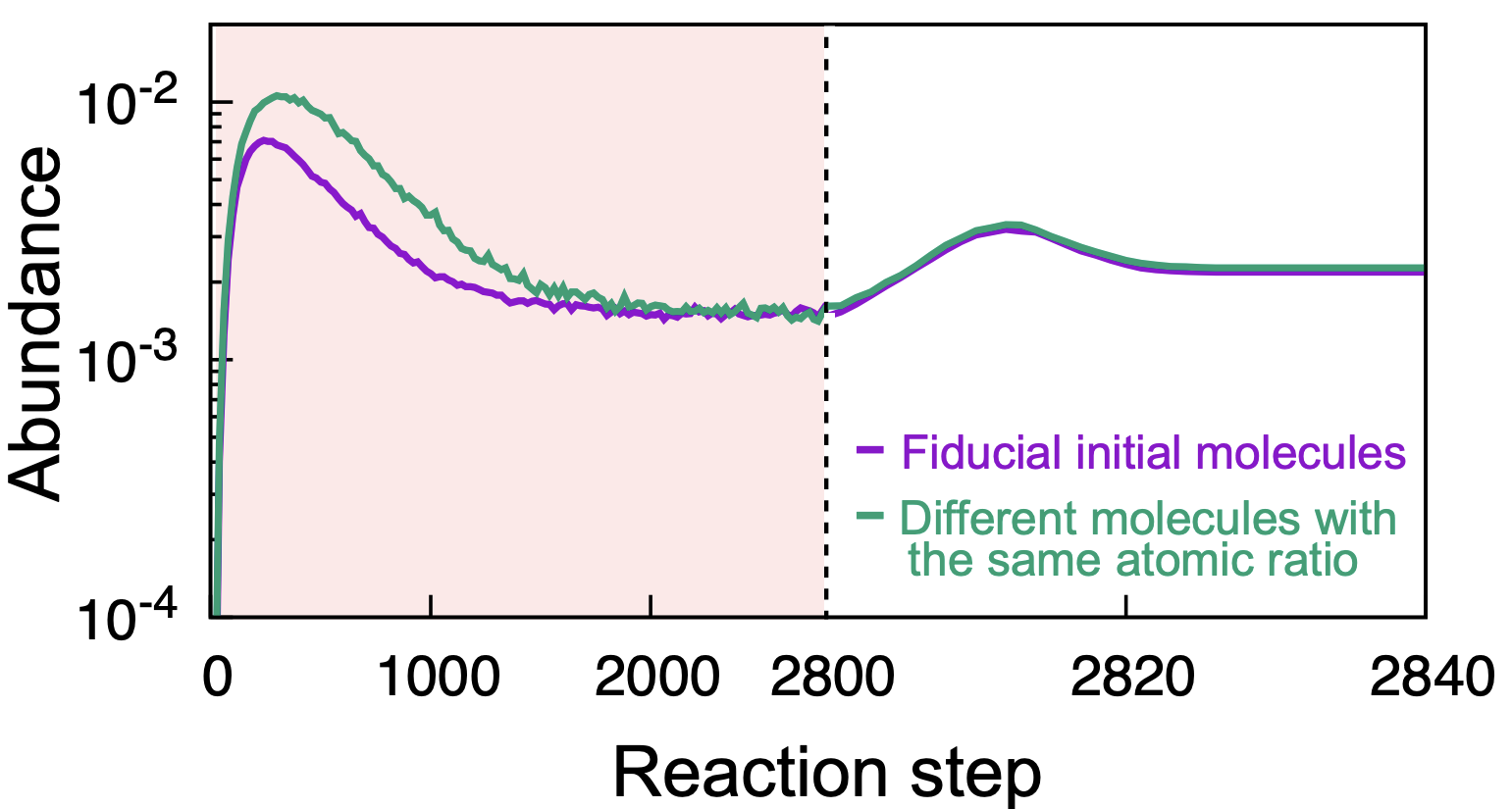}
  \caption{
  Change in the abundances of amino acids obtained with the fiducial initial molecules (magenta curve) and the initial molecules with the same atomic ratio but different molecular species (green curve), which consist of 7 carbon dioxide molecules ($\rm CO_2$), 9 ammonia molecules ($\rm NH_3$), 15 water molecules ($\rm H_2O$), and 16 hydrogen molecules ($\rm H_2$).
  While the initial evolution, which depends on the initial molecular species, differs between the two, they synchronize once the bonds are sufficiently randomized by UV irradiation.}
  \label{fig:diff_mol}
\end{figure}

\begin{figure}
  \centering
  \includegraphics[width= 7cm]{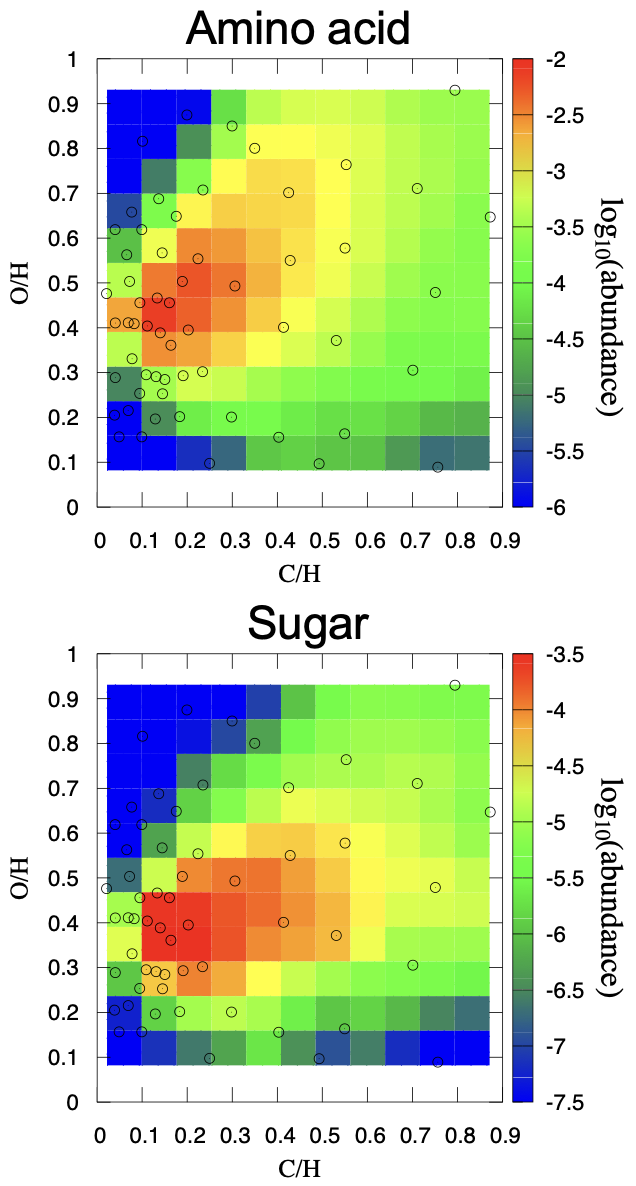}
  \caption{Dependence of the final abundances of amino acids (upper panel) and sugars (lower panel) on the initial atomic ratios of C/H and O/H. The small open circles represent the parameters with which we performed the simulations (Table~\ref{tab:initialsets}). The heat maps are created with $\log_{10}(\rm abundance)$. The color level of sugars is 1.5 orders lower than amino acids (see discussions in Sect.~\ref{sec:semi_analy}) to compare the dependence patterns on O/H and C/H between amino acids and sugars.
  } 
  \label{fig:contour}
\end{figure}


In this section, we examine how the final abundances of amino acids and sugars depend on the initial atomic ratios.
Differences in initial molecular species are not considered here since they produce 
only minor variations, while differences in atomic ratios can change the abundances of amino acids and sugars by several orders of magnitude.

The 57 sets of initial molecules used for the survey are summarized in Table~\ref{tab:initialsets}. 
The molecular species were mainly chosen from ones discovered in interstellar ice \citep[e.g.,][and references therein]{Gibb2004-ei}, while also including species not currently detected.
We would like to emphasize that in our calculations, the molecular distribution after undergoing sufficiently long UV irradiation is determined by the atomic ratios of initial molecules, rather than the molecular species.
Figure~\ref{fig:diff_mol} shows the change in the abundance of amino acids obtained from the initial molecules where carbon molecules are replaced with $\rm CO_2$, while maintaining the same atomic ratio as the fiducial set.
The results in Fig.~\ref{fig:diff_mol} indicate little dependence on the initial molecular species.
Thus, the molecular sets in Table~\ref{tab:initialsets} do not necessarily represent realistic compositions, but were designed to provide various ratios of C/H and O/H.
They are also designed so that the total number of carbon, nitrogen, and oxygen atoms is 40-55, assuming that a similar number of atoms can interact in the same conditions, while the absolute number is arbitrary.
Since the dependence of amino acid and sugar synthesis on N/H ratio is relatively weak, the N/H ratio of these molecular sets was fixed at 0.1.

Heat maps of final abundances of amino acids and sugars created from results from these initial molecular sets are shown in Fig.~\ref{fig:contour}.
It shows that slight differences with $\sim 0.1$ in the initial atomic ratios, C/H and O/H, can produce a several orders of magnitude variation in the final abundances.

As already shown in Figs.~\ref{fig:p3aminosugar} and ~\ref{fig:p44aminosugar}, 
the abundance evolution of amino acid and sugar are remarkably similar. Figure~\ref{fig:contour} shows that there is a quite similar dependence of the final abundance between them on C/H and O/H, as well.
The peaks are at C/H $\simeq 0.15$ and O/H $\simeq 0.4$-0.5 for amino acids, and C/H $\simeq 0.15$ and O/H $\simeq 0.3$-0.4 for sugars, and the abundance magnitude is about 1.5 orders higher for amino acids than for sugars in the entire parameter space that we have examined.

Each of the 57 calculation results in Fig.~\ref{fig:contour} was obtained through $10^6$ runs generating reaction sequences with different sets of random numbers.
We would like to stress that the scheme described in Sect.~\ref{sec:method} enables us to carry out such a broad parameter survey even for low-yield products and to capture the overall features of the synthesis of amino acids and sugars.
This cannot be addressed with experiments and quantum chemistry simulations at present and would give insights into the intrinsic synthesis mechanisms.

\subsubsection{Introduction of an index of molecular complexity} \label{sec:complex}

\begin{figure*}
  \centering
  \includegraphics[width=15cm]{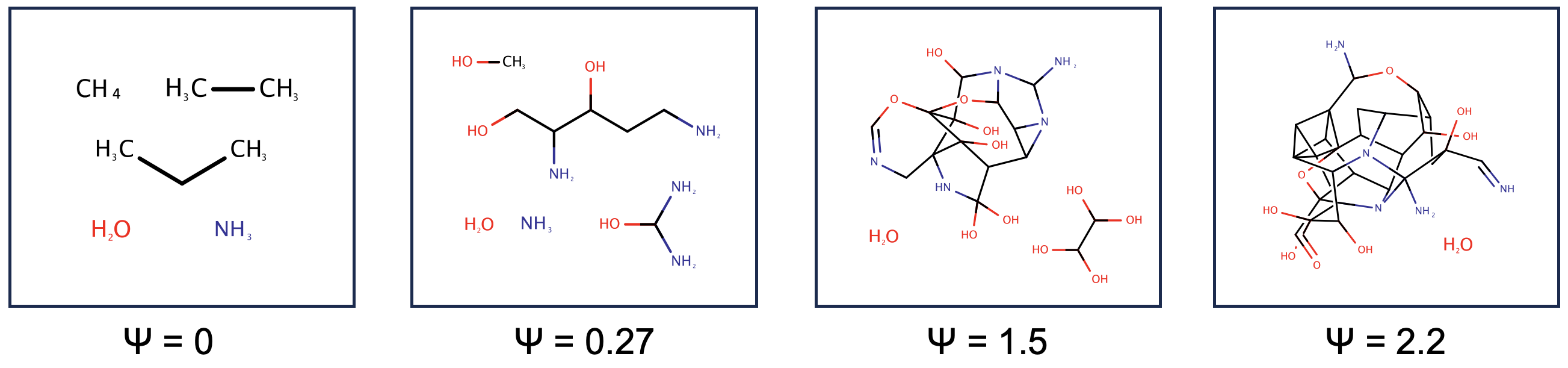}
  \caption{Typical end products of the initial molecular sets No.3, 1, 6, and 44, which have $\Psi \sim 0, 0.27, 1.5$, and 2.2, respectively.
  These products are obtained by the Monte Carlo simulations.}
  \label{fig:final_products}
\end{figure*}

\begin{figure*}
  \centering
  \includegraphics[width=14cm]{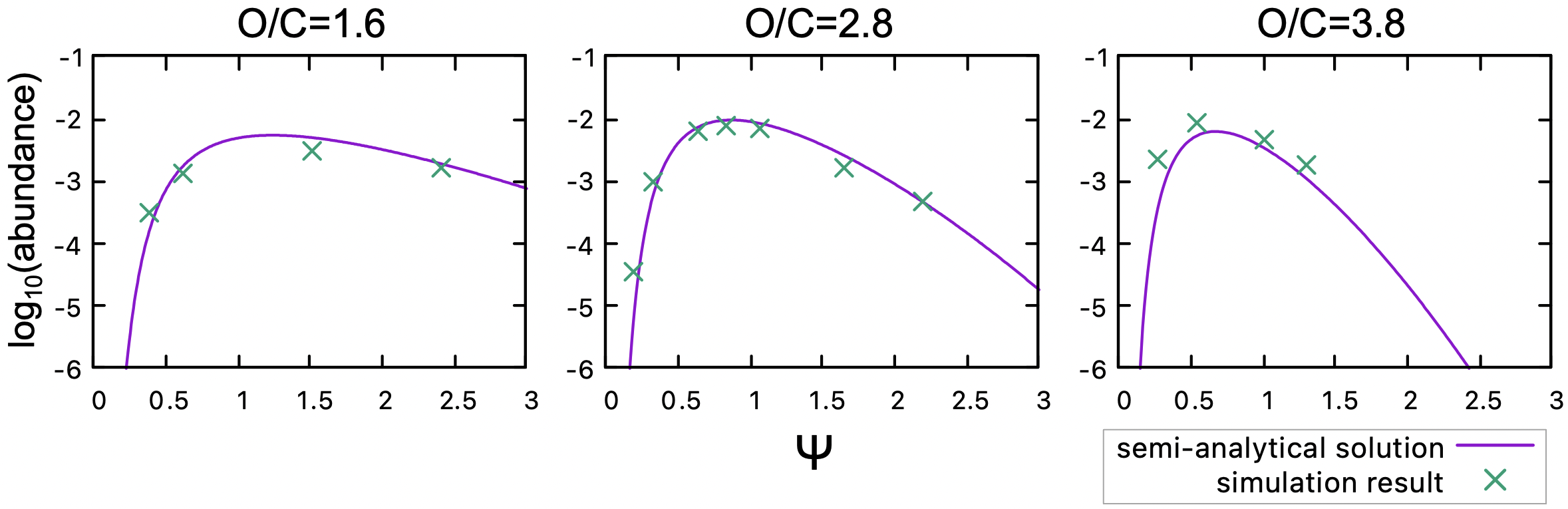}
  \caption{Final abundance distribution of amino acids as a function of $\Psi$ with $\rm O/C \sim$ 1.6, 2.8, and 3.8. The result of the Monte Carlo simulations is plotted by a green x mark, and the prediction by the semi-analytical formula (Eq.~(\ref{eq:semiAS})) is plotted by a magenta curve. The O/C ratios of the molecular sets used in the plot contain a maximum error margin of about 9\%.
  }
  \label{fig:Psi_dep}
\end{figure*}

To understand the dependence of the final abundance, we introduce an index for the potential complexity of the molecular set as follows:
\begin{equation}
    \label{eq:Psi}
    \Psi =\rm{\frac{4C+3N+2O-H}{H}}.
\end{equation}
Here, C, N, O, and H represent the number of carbon, nitrogen, oxygen, and hydrogen atoms in the initial molecules, respectively. 
The coefficients for C, N, and O represent valencies (the number of unpaired electrons) for each.
Assuming that all the hydrogen atoms are bonded to C, N, and/or O, this number means the total number of remaining unpaired electrons in the non-hydrogen atoms (C, N, and O) relative to the number of bonds to the hydrogen atoms.

For example, the collection of methane, ammonia, and water molecules gives $\Psi = 0$. 
The products of those molecules mostly go back to the initial H-saturated molecules in the post UV phase, providing simple end products as shown in the leftmost panel of Fig.~\ref{fig:final_products}. 
In such conditions, complex organic molecules containing O and N atoms, such as amino acids and sugars, are eventually broken down
\footnote{Considering highly H-rich case, such as the molecular sets with large fraction of H$_2$, $\Psi$ can be negative. As with the example of $\Psi =0$, all final products are saturated with H atoms, giving no differences from $\Psi =0$ case in the complexity of carbon products. Therefore, $\Psi<0$ is given $\Psi =0$.}.

In the fiducial set with C = 7, N = 9, O = 29, and H = 89, Eq.~(\ref{eq:Psi}) gives $\Psi \rm H = 24$ ($\Psi \simeq 0.27$), meaning that even if all the hydrogen atoms are bonded to the C, N, or O atoms, 24 unpaired electrons remain among those atoms.
Those unpaired electrons are potentially able to form other types of bonds such as C-O, C-N, and N-O.
Typical final products in the fiducial set obtained by the Monte Carlo simulation are shown in the second panel from the left in Fig.~\ref{fig:final_products}. 
In this manner, larger $\Psi$ results in the formation of more chemical bonds among C, N, and O atoms, leading to the increase in molecular complexity and diversity in general.

However, when $\Psi$ becmoes larger, for example, $\Psi \ga 2$, final products are dominated by C-C and/or C-O bonds, and the molecules are generally large and complex.
Typical final products for $\Psi \sim 2$ are shown in the rightmost panel of Fig.~\ref{fig:final_products}. 
In this case, contrary to the situation when $\Psi =0$, C-H bonds are depleted, resulting in small amounts of amino acids and sugars.

The final abundances of amino acids obtained by the Monte Carlo simulations with $\rm O/C \sim$ 1.6, 2.8, and 3.8 are plotted as a function of $\Psi$ in Fig.~\ref{fig:Psi_dep}.
As explained above, they have peaks when $\Psi$ is moderate:
while it depends on O/C ratio, the typical value of $\Psi$ providing the peak was found to be around 0.5-1.0.


\subsubsection{Semi-analytical formula to predict final abundances of amino acids and sugars from initial C/H and O/H ratios} \label{sec:semi_analy}

\begin{figure}
  \centering
  \includegraphics[width= 7cm]{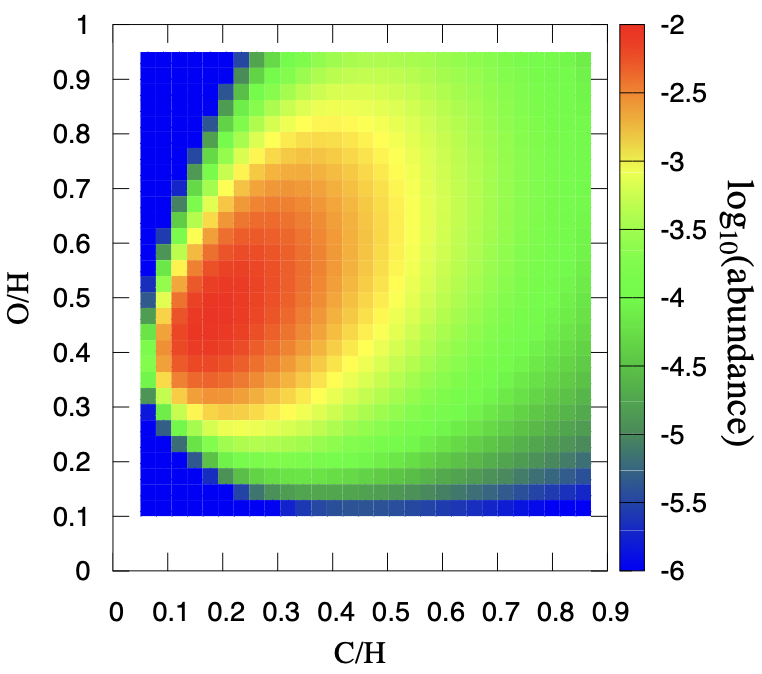}
  \caption{Final abundance distribution of amino acids on the C/H-O/H plane, predicted by the semi-analytical formula (Eq.~(\ref{eq:semiAS})).
  The heat map is created with $\log_{10}(A_{\rm final})$.
  }
  \label{fig:analytical}
\end{figure}

Using the index $\Psi$ and the initial atomic ratios, 
a semi-analytical formula for the final abundances of amino acids and sugars ($A_{\rm fin,anly}$) is derived as follows:
\begin{equation}
    \label{eq:semiAS}
    A_{\rm fin,anly} = c_1 \bigg(\frac{\rm O}{\rm H}\bigg) \bigg(\frac{\rm O}{\rm C}\bigg) ^2 \exp \biggl\{-c_2\frac{1}{\Psi}\frac{\rm H}{\rm O} - c_3\Psi - c_4 \bigg( \frac{\rm O}{\rm H} \bigg)^2\bigg( \frac{\rm O}{\rm C}\bigg)^2 \biggr\},
\end{equation}
where $c_2, c_3$ and $c_4$ are constants of roughly $\sim O(1)$ and $c_1$ is an arbitrary scaling parameter, which are determined by the structural properties of the individual molecules and depends on the simulation parameters. 
In this study, their specific values were obtained by fitting them to the numerical results.
We adopt the scaling parameter of $c_1 = 0.1$ and the constants of $O(1)$ as $c_2 = 0.5, c_3 = 1.4$ and $c_4 = 0.7$ as the fitting values for an amino acid abundance.
The contour plot and the cross-section plots of Eq.~(\ref{eq:semiAS}) for amino acids are shown in Figs.~\ref{fig:analytical} and ~\ref{fig:Psi_dep}, respectively.
Equation~(\ref{eq:semiAS}) reproduces the results from Monte Carlo calculations over a wide range of initial atomic ratios.

Each negative exponent term in Eq.~(\ref{eq:semiAS}) expresses a mechanism that inhibits the amino acid and sugar formation, as explained below.
Thus, understanding the structure of this semi-analytical formula would be useful for comprehending the favorable atomic ratios for the synthesis of amino acids and sugars.

Under the condition dominated by random bond rearrangements in the UV phase as described above, yields of each product are controlled by the abundances of the constituent bonds.
Hence, the final abundance of amino acids is also determined by the final bond ratios of the molecular set.
Based on the structures of amino acids and sugars,
four different mechanisms 
limit their final abundance: the depletion of 1) C=O relative to C-H, 2) C=O relative to C-C, 3) C-H relative to C-C, and 4) C-H relative to C-O in the final bond distribution.


The first factor in the exponential function in Eq.~(\ref{eq:semiAS}) represents the depletion of C=O relative to C-H when $\Psi \ll 1$, corresponding to excess of H atoms.
In such a condition, the bonds between non-hydrogen atoms are less likely to form. 
Since the formation of C=O bonds has the lowest abundance among the constituent bonds in a broad range of C/H and O/H ratios, including the fiducial case (see the left panel of Fig.~\ref{fig:bondchange}), the production of C=O serves the rate-limiting factor for the synthesis of amino acids and sugars when $\Psi \ll 1$.
In this case, a large H/O ratio is detrimental to the C=O formation, because O atoms are deprived of H atoms, which prevents an adequate supply of O atoms to the carbon atoms to form C=O bonds.
This suppression mechanism is therefore expressed by $\exp \big(-(1/\Psi)(\rm H/O) \big)$.

Depletion of C=O bonds is also brought about by an increase in C-C bonds within the carbon bond budget when the C/O and H/O ratios of the molecular set are too large.
These effects are represented by the pre-exponential factor $\big(\rm O/H\big) \big(\rm O/C\big)^2$.
Importantly, this factor also reflects the increasing trend in their abundances accompanying the rise in the O/C ratio from zero.
Since C=O formation is a bottleneck in the synthesis of amino acids and sugars over a wide range as described above, enhancing the O/C and O/H ratios facilitates
amino acid and sugar synthesis by increasing the C=O bonds.

While the increase in $\Psi$ near zero gives rise to the formation of amino acids and sugars, further increase of $\Psi$ beyond $\sim 1$ leads to the depletion of C-H relative to C-C, which is disadvantageous to the formation of amino acids and sugars.
The final products of the molecular set with large $\Psi$ tend to be too complex molecules, as explained in Sect.~\ref{sec:complex}.
The third exponent term $\exp (-\Psi)$ represents the decrease in amino acids and sugars due to too low C-H/C-C ratio.
This factor combined with the first one, $\exp \big(-(c_2/\Psi)(\rm H/O) -c_3 \Psi \big)$ indicates the preference of amino acids and sugars for intermediate $\Psi$.

The synthesis of amino acids and sugars is also inhibited when C-H is depleted relative to C-O.
As in the former case, depletion of C-H bonds decreases amino acid and sugar production.
If O/H and O/C ratios are very large, excess O atoms deplete C-H bonds by depriving H atoms from C atoms to form stable O-H bonds.
At the same time, large O/H and O/C ratios produce abundant C-O bonds, resulting in C-O bond occupancy in the carbon bond budget.
Hence, this inhibitory mechanism is represented as $\exp \big(-(\rm O/H)^2 (O/C)^2\big)$.

We emphasize that the coefficients, $c_i \: (i=1$-4), and functional forms used in Eq.~(\ref{eq:semiAS}) are chosen to fit with data and are arbitrary at this moment.
Nevertheless, as already mentioned, the construction of this semi-analytical formula demonstrates the important mechanisms behind the synthesis of amino acids and sugars, depending on the initial atomic ratio.

The coefficients may change when the N/H ratio and other parameters change.
However, the dependence on N/H is weak, as is suggested by the fact that it is not involved in the decreasing and increasing factors of amino acids and sugars. 
As shown in Fig.~\ref{fig:group}, the amino group, which is another important part of amino acids, is abundant  in general, unless the N/H ratio is closer to zero.
Even with N/H = 0, our simulation obtains a similar abundance pattern of sugars on the C/H-O/H plane, although the absolute values of the abundance are enhanced. 
On the other hand, for a very large N/H ratio, sugar formation is more inhibited overall.
A more general formulation is left for a future work.

\subsection{Dependence on temperature}\label{sec:d_tem}

\begin{figure}
  \centering
  \includegraphics[width=7cm]{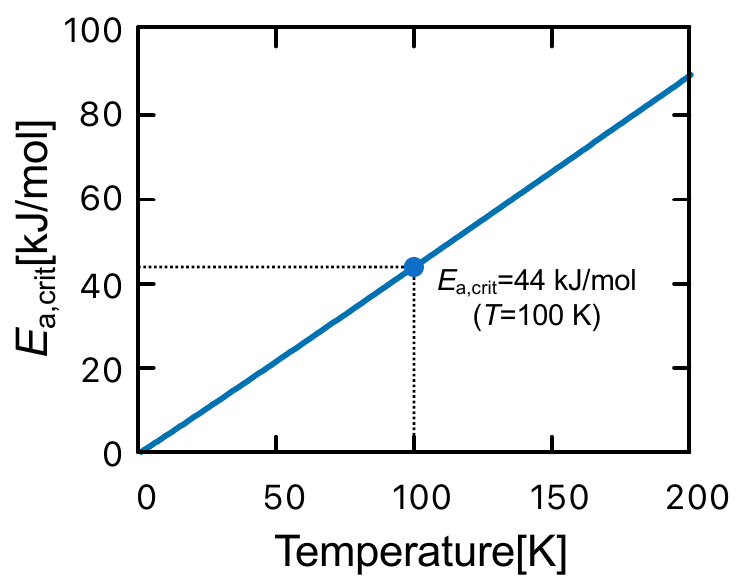}
  \caption{Relation between $E_{\rm a,crit}$ and temperature $T$ based on the Eyring equation (Eq.~(\ref{eq:Ecrit})) with $t_{1/2}=$1000 years.}
  \label{fig:Ecrit_temperature}
\end{figure}

\begin{figure}
  \centering
  \includegraphics[width=8cm]{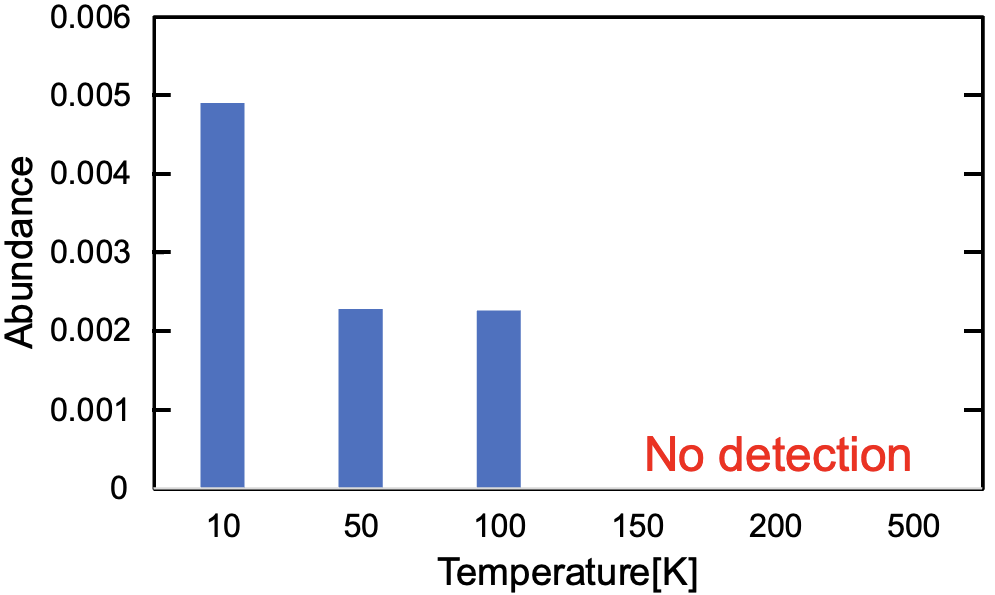}
  \caption{Dependence of the final abundance of amino acids on temperature, $T$. The other parameters are the same as the fiducial set.}
  \label{fig:d_tem}
\end{figure}

Dependence of the final abundances of amino acids and sugars on the temperature is shown in Fig.~\ref{fig:d_tem}.
Since temperature is a parameter of the weighting equation (Eq.~(\ref{eq:W-weight})) and the Eyring equation (Eq.~(\ref{eq:Ecrit})), using different temperatures changes relative probabilities between the candidate reactions and the magnitude of $E_{\rm a,crit}$ (Fig.\ref{fig:Ecrit_temperature}).
In other words, as temperature rises, the dependence of the reactions on temperature becomes weaker, and the reaction with higher activation energy proceeds within the same timescale.

Those changes have no effect on the reactions during the UV phase in the temperature range examined here since only barrierless reactions occur in this phase.

On the other hand, the reactions in the post-UV phase strongly depend on $E_{\rm a,crit}$.
When $T \lesssim  30\,\rm K$, even after the UV phase finishes, almost only barrierless reactions proceed due to low $E_{\rm a,crit}$.
Hence, the reactions are limited to radical-radical reactions and some barrieless type 3 and 4 reactions.
As temperature increases, reactions with activation energies are allowed and start to compete with barrierless reactions.

Figure~\ref{fig:d_tem} shows the final abundance of amino acids at temperatures ranging from 10 K to 200 K in the fiducial set of initial molecules. 
The abundance decreases at $T$ between 10 and 50 K and completely diminishes between 100 and 150 K. 
In this molecular set, the reactions i) C-O + H-H $\rightarrow$ C-H + O-H with $\Delta H = -84 \, \rm kJ/mol$ (Eq.~(\ref{eq:COHH})) and ii) C-O + C-H $\rightarrow$ C-C + O-H with $\Delta H=-39 \, \rm kJ/mol$ (Eq.~(\ref{eq:COCH})) start to occur at $T \gtrsim 40 \, \rm K$ and $T\gtrsim 140 \, \rm K$, respectively, because the activation energies of these reactions are $E_{\rm a}= 16 \, \rm kJ/mol$ and $E_{\rm a}= 61 \, \rm kJ/mol$ (Eq.~(\ref{eq:evans2})) while the critical activation energy is given as $E_{\rm a,crit} \simeq 44 \,(T/100\, \rm K) \, \rm kJ/mol$ (Eq.~(\ref{eq:Ecrit})).
These reactions decompose C-O bonds and C-H bonds, leading to the decomposition of amino acids and sugars accordingly. 
As a result, the abundance of sugars exhibits a similar temperature dependence.
Figure~\ref{fig:d_tem} shows the reaction ii) is more effective to decompose amino acids.

This trend that the abundances of amino acids and sugars decrease in higher temperatures may be counter-intuitive
since build-up-type reactions such as the Strecker reaction require thermal energy \citep{Magrino2021}.
Since the synthetic process identified in our simulations is a photon-driven reaction by UV irradiation, 
which is a non-thermal process, low temperature conditions enhance the final abundances of COMs by protecting the products from decomposition.


While lower temperatures are favorable for preservation of amino acids and sugars, 
no significant 
difference in the final abundances were observed over a wide temperature range of $T=\rm \, 40$-130 K, including the fiducial parameter of 100 K, because there is no key reactions for the decomposition that are switched in this temperature range.
This means that the results shown in previous sections do not significantly depend on the assumed ice temperature in a range of $T \sim 40$-130 K.

We note that the current simulation assumes that 
molecular diffusion on an icy grain surface is efficient enough for all the molecules within the set to react with each other. 
The temperature range that is actually applicable for this assumption can be 
narrow because the molecular diffusion would be inefficient at low temperature.
The issue regarding diffusion is left for future work.

We also note that the transition temperature, $T\sim 130 \, \rm K,$ depends on the choice of $\beta$ in the Bell-Evans-Polanyi principle (Eq.~(\ref{eq:evans2})).
As we discuss in Sect.~\ref{sec:d_ab}, 
$\beta$ could be higher than the fiducial value of $\beta = 100 \,\rm kJ/mol$ up to $\beta \sim 180 \, \rm kJ/mol$, depending the types of reactions.
If $\beta = 160 \, \rm kJ/mol$ is adopted,
the reaction ii) gives $E_{\rm a}= 121 \, \rm kJ/mol$, and the critical temperature becomes 275 K.

\subsection{Dependence on photon energy} \label{sec:d_photonE}

\begin{figure}  \centering  
\includegraphics[width=8cm]{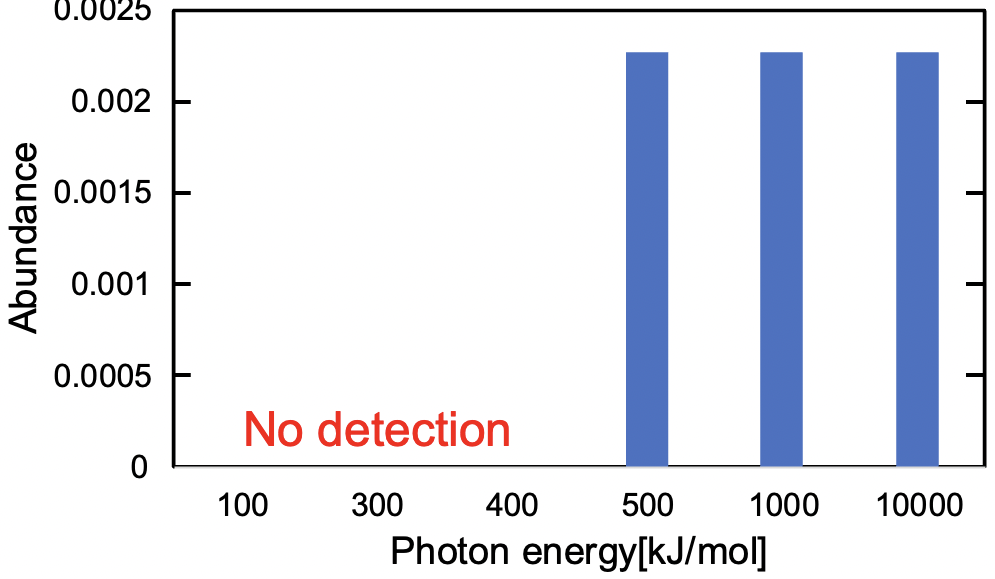}  
    \caption{Dependence of the final abundance of amino acids on photon energy. The other parameters are the same as the fiducial set. 
    }  
    \label{fig:d_photonE} 
\end{figure}

In our calculations, photon energy is converted to the negative bond energy of a hypothetical molecule $\rm{X_2}$ (see~Sect.~\ref{sec:radicals}). The final abundances of amino acids obtained with different $\rm{X_2}$ bond energies are shown in Fig.~\ref{fig:d_photonE}.

Figure ~\ref{fig:d_photonE} shows that the abundance rapidly diminishes as the photon energy changes from $- 500 \, \rm kJ/mol$ to $- 400 \, \rm kJ/mol$.
This drastic transition is caused by the cessation of photodissociation of water molecule.
When the $\rm{X_2}$ bond energy is larger than $\sim - 464 \, \rm kJ/mol$, meaning that the photon energy is lower than $464 \, \rm kJ/mol \simeq 4.6 \, \rm eV$, $\rm H_2O$ molecules in the initial molecules cannot become radicals.
As a result, sufficient supply of oxygen atoms to carbon atoms does not occur, and the formation of functional groups is inhibited.
On the other hand, if the $\rm{X_2}$ bond energy is $\lesssim - 464 \, \rm kJ/mol$, all the bonds in this simulation are photodissociated with equal probabilities.
The reaction process is the same as the fiducial case in this range.

Since the critical photon energy causing a drastic change depends on the initial molecular species and on the bond energies used, we do not focus on the absolute value of this threshold.
To accurately evaluate this boundary, it is necessary to consider the photoabsorption cross section for the target UV wavelength for each species.

\subsection{Dependence on $\alpha$ and $\beta$ of the Bell-Evans-Polanyi principle} \label{sec:d_ab}

\begin{figure}
  \centering
  \includegraphics[width=7.5cm]{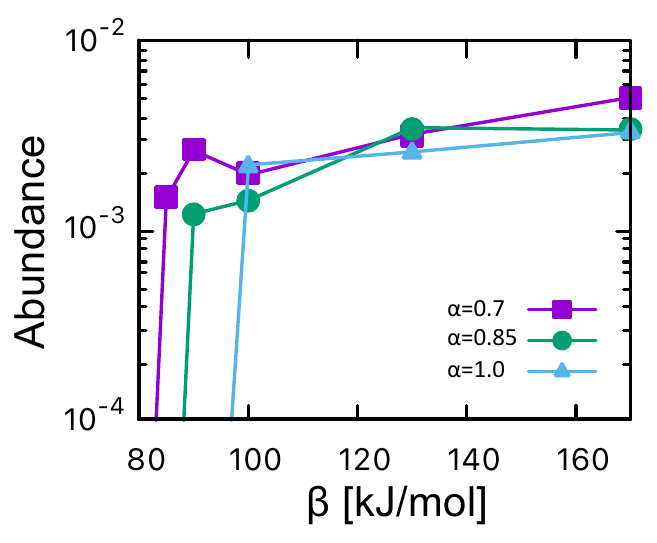}
  \caption {Dependence of the final abundance of amino acids on the $\alpha$ and $\beta$ of the Bell-Evans-Polanyi principle. The results were calculated in the range of $\alpha=0.7-1.0$ and $\beta=80-170\rm\, kJ/mol$ at $T=100\, \rm K$. The other parameters are the same as the fiducial set.}
  \label{fig:d_ab}
\end{figure}

The Bell-Evans-Polanyi principle, which gives the approximate relationship between an enthalpy change and an activation energy, contains two parameters, $\alpha$ and $\beta$ (see Sect.~\ref{sec:BEP}).
In our calculations, $\alpha$ and $\beta$ affect the $\Delta H$ value giving $E_{\rm{a}}=0$ and the $\Delta H$ value giving $E_{\mathrm{a}}=E_{\rm{a,crit}}$. 
Here, $\alpha$ also changes the relative probabilities between the candidate reactions.
As pointed out in Sect.~\ref{sec:BEP}, while the Bell-Evans-Polanyi principle is simple and useful, the parameter values have uncertainty.
The density function theory calculations 
\citep{michaelides_2003,Wang_2011_evans,sutton_2012}
show that while $\alpha$ is usually $\sim 1$ or slightly smaller, $\beta$ could be bimodal; $\beta \sim 170-180 \, \rm kJ/mol$ or more for reactions cleaving bonds between C, N, and O atoms, and $\beta \sim 80-100\, \rm kJ/mol$ for reactions cutting C-H.
However, there are exceptions and it is not clear if the same trends exist for bonds in COMs.

Nonetheless, since during the UV phase, reactions are dominated by barrierless radical-radical reactions and photodissociation reactions, the uncertainty does not affect the results.
Therefore, the uncertainty in $\alpha$ and $\beta$ essentially influences only the reactions in the post-UV phase.

Figure~\ref{fig:d_ab} shows the final abundances of amino acids and sugars using different pairs of $\alpha$ and $\beta$ for $T=100\, \rm K$. 
In the range predicted by DFT calculations for $\alpha$ and $\beta$, most cases show the similar abundances, with differences of only a few times.

However, when $\beta \simeq 90 \,\rm kJ/mol$, the abundances drastically drop down.
This is because $E_{\rm a}$ of the reaction: C-O + C-H $\rightarrow$ C-C + O-H $ (\Delta H = -39 \rm \, kJ/mol)$ is evaluated as about 53 kJ/mol, 47 kJ/mol and 41 kJ/mol with $\alpha$=0.7, 0.85, and 1.0 all with $\beta$=80 kJ/mol, respectively.
Since $E_{\rm a,crit}$ is set to 44 kJ/mol, 
it is predicted that
hydroxyl group (-OH) and C-H bonds necessary for amino acids and sugars efficiently decompose in these ranges.

On the other hand, these results show that the dependence of the final products on $\alpha$ and $\beta$ is caused solely by whether particular reactions are allowed to proceed in the post-UV phase according to $E_{\rm a,crit}$.
In other words, most reaction processes are weakly dependent on $\alpha$ and $\beta$, and this dependence is important only when predicting the preservation process of the final products.
For $\alpha = 1$, the activation energy for the decomposition reaction: C-O + C-H $\rightarrow$ C-C + O-H $ (\Delta H = -39 \rm \, kJ/mol)$ is given as $E_{\rm a} = \beta -39 \rm \, kJ/mol$, based on Eq.~(\ref{eq:evans}).
Therefore, the larger the $\beta$ value, the larger the evaluated activation energy of the decomposition reaction.
As mentioned in Sect.~\ref{sec:d_tem}, larger $\beta$ such as $\beta = 160 \, \rm kJ/mol$ could prevent the decomposition process from rapidly occurring even at room temperature.



\subsection{Dependence on timescale}\label{sec:d_time}

Since the reaction steps in our calculation are not associated with a specific time length, the simulation timescale is constrained by $E_{\mathrm{a,crit}}$ (see Sect.~\ref{sec:t_Ea}).

As only barrierless reactions happen during the UV phase at low temperatures, the change in $E_{\mathrm{a,crit}}$ does not impact the UV phase reactions.
Since the current simulation does not take into account composition changes, the reactions in the UV phase are calculated only to examine the changes after the UV irradiation is turned off (see Sect.~\ref{sec:meaningofIM}).
If we consider those processes, the dependence on the timescale, especially the composition change by desorption in a long timescale, could be significant.

Whereas the activation energy of the reaction occurring in the post-UV phase is controlled by $E_{\mathrm{a,crit}}$, the abundances did not show any significant changes in the range of $\Delta t=1-10^6 \, \rm{years}$ either.
This is because the dependency of $E_{\mathrm{a,crit}}$ on the timescale is too small to induce changes in the occurring reactions (Eq.~(\ref{eq:Ecrit})).
Therefore, a wide variety of organic molecules, including biomolecules could be preserved if the ice grains are incorporated in small bodies before the molecules are desorbed from the surface or altered at high temperatures.

\section{Discussion}
\label{sec:discussion}
\subsection{Challenges regarding comparison with experiments and observations}
\label{sec:experiments}

Our goal for developing the new Monte Carlo simulation is to construct a theoretical model that allows for direct comparison with experiments and observations.
This would lead to the interpretation of the experimental and observational data, and to the proposal of new experiments and observations as working hypotheses.
At the same time, experimental and observational data are used to calibrate the model through comparison.
The present study is a first step to connect theory, experiments, and observations, aiming to enhance our comprehension of COM synthesis in space.
In order to achieve this purpose, more updates are needed both in the theoretical model and the experiments/observations, as discussed below.


In ice irradiation experiments, it is generally challenging to identify complex molecules, such as amino acids and sugars, in situ. 
For detailed analysis, such as mass spectrometer and chromatography analysis, the irradiated ice samples are usually warmed up to room temperature.
The volatile elements sublimate from the sample during this process, and the refractory residues are used for the analysis of complex organic compounds.
In the amino acid analysis, the organic residues are decomposed with acid hydrolysis to conduct liquid chromatography analysis. 
It is well known that this hydrolysis process produces abundant amino acids from the residues \citep[e.g.][]{Nuevo2008}.
The relationship between the amino acids predicted to form in ice based on our results and those detected through the hydrolysis of refractory samples is not clear at the moment.

Our results suggest that amino acids and sugars decompose at a temperature above 140 K for the fiducial $\beta = 100 \, \rm kJ/mol$. 
Even if the limit timescale is changed from $10^3$ years to $10^{-3}$ years ($\sim 10$ hours), it is predicted that they immediately decompose at room temperature due to the strong dependence of the chemical reactions on temperature.
It may be consistent with the experimental fact that the amount of free amino acids eluted from the residue is significantly small. 
For comparison with the amino acids appearing after hydrolysis, our simulation needs to take the post-experiment processes into account because those processes can alter the original ice products through 
desorption of volatile species, thermal promotion of chemical reactions in the residue and the hydrolysis.
Since the hydrolysis process would have also occurred in the parent bodies of carbonaceous chondrites, this effect would be exceedingly important for the chemical evolution in protoplanetary disks.

On the other hand, as we pointed out in Sects.~\ref{sec:d_tem} and ~\ref{sec:d_ab}, if $\beta$ is close to the upper limit of the theoretical prediction ($\sim 170-180 \, \rm kJ/mol$), the decomposition of amino acids at room temperature can be marginal. It may be more consistent with the empirical fact that free amino acids are detected at room temperature although the abundance is relatively low and that the amino acids formed by hydrolysis are not immediately decomposed.



Sugars are also detected from the UV-irradiated ice samples in ex-situ analysis at room temperature \citep[e.g.,][]{Meinert2016,Marcellus2015}.
While our current simulation does not take the effect of stereo-structure into account, sugars are stabilized when they take the ring structures \citep[e.g.,][]{Azofra2012,Dass2021}.
As discussed in \citet{takehara}, this stabilization of sugar molecules could prevent them from decomposition.
Our results on temperature dependence suggest that the degradation reactions of sugars and amino acids progress rapidly with relatively small changes in $E_{\rm a,crit}$. 
If the activation energy of the decomposition reaction is about 20 kJ/mol higher than the present calculation, the progress of the sugar decomposition can be avoided.
If $\beta$ is higher than the fiducial value ($100 \, \rm kJ/mol$), the stabilization by transformation to the ring structure would be more robust.


Just as our simulation needs to be updated, there is a strong need for a more thorough  investigation of photoreactions in ice through in-situ analysis.
Especially, quantitative experimental data focusing on the dependence on the initial atomic ratios, a factor suggested to be crucial in determining the final products in this study, is highly desired.
The in-situ analysis of cometary samples \citep[e.g.,][]{Hanni2020, Goesmann2015} and JWST observations of interstellar icy dust \citep[e.g.,][]{McClure2023,Rocha2024} would also allows for the direct comparison between the theoretical prediction and realistic outcomes.
While free amino acids are hardly found in the ice irradiation experiments as described above, glycine, the simplest amino acid, has been detected in cometary samples \citep{Altwegg2016}.
The presence of amino acids in a primitive icy object may be consistent with the formation and preservation of amino acids at low temperatures predicted by our calculations.

 

It is also important to consider the conditions under which our results are applicable, based on the approximations and assumptions used in our calculations.
As already pointed out, the higher the temperature, in other words, the greater the influence of desorption, the more significant the deviation from this simulation which assumes a closed system.
Even in such a case, as described in Sect.~\ref{sec:meaningofIM}, as long as enough randomization of bonds occurs in ice during UV irradiation, the final products would show the dependence on the atomic ratios just before the UV irradiation stops, which were found in the present calculations.

Our simulations also assume that photodissociation and molecular diffusion are adequately active, and their dependence on the species is negligible.
The dependence on diffusion, qualitatively speaking, suggests that molecules with smaller masses tend to diffuse more easily \citep[e.g.,][]{garrod_herbst2006,chang_2016}.
This may make the production of larger molecules more inefficient.
The cage effect might also hinder the formation of complex molecules as the radicals can interact only with adjacent molecules and radicals.
In the case where the dependence of photodissociation on the molecules and/or the bond-type is strong, selective deconstruction and/or production of particular molecules should take place; this may result in reaction networks different from the present ones.
However, \citet{Henderson2015} referred to the similarity between the products in ultraviolet (Ly$\alpha$) and electron-irradiation which is basically able to break any type of bond.
Increasing molecular complexity may lead to averaging of properties from molecule to molecule. 
This may support the validity of our assumption that all bonds can be dissociated by UV irradiation.

Regarding the temperature dependence, inhibition of molecular desorption at low temperatures may promote the synthesis of some organic species.
On the other hand, extremely low temperatures negatively work for COM synthesis because molecular diffusion becomes inefficient \citep{Tsuge2023}.
These effects compete in reality, making it difficult to comprehend the reaction process behind the COMs' formation.
In this study, we found that the lower temperatures result in a higher yield of COMs if we consider purely the reaction network, neglecting the thermal motion of molecules (Sect. \ref{sec:d_tem}).

\subsection{Implications for COM synthesis in protoplanetary disks}
\label{sec:IOM}
Organic syntheses in protoplanetary disks should be contributed to several processes, such as the gas phase reactions in the photosphere of the disk \citep[e.g.,][]{Kuga2015, Bekaert2018}, the liquid phase reactions in asteroids (parent bodies of chondrites) by heat \citep{Koga2022} and/or gamma-rays caused by radiogenic $^{26}$Al \citep{Kebukawa2022}, and the ice surface reactions in the outer region of the disk \citep[e.g.,][]{Oberg2009} (also see \citet{Jorgensen2020} for a review).

Our results demonstrate that the radical-dominated reactions driven by UV irradiation on ice surfaces could provide diverse and complex organic molecules due to the randomization of the covalent bonds.
Therefore, the photochemistry occurring in icy dust grains could be a very favorable mechanism for the synthesis of COMs including amino acids, sugars, and arguably other biomolecules in protoplanetary disks.
In this work, we evaluated the production of amino acids and sugars using the abundance defined by Eq.~(\ref{eq:abundance}). 
It should be noted that converting the abundance used in this study to the absolute yield in realistic ice mantles requires careful consideration.
In addition, the detectability of molecular species depends on the ability of telescope as well as the yield.
So far, even if amino acids and sugars present in the disk, the detection is challenging with current observational techniques, especially in optically thick disks.
Although our work predicts the production of amino acids and sugars in protoplanetary disks, further studies are required to determine whether they can be observed, which is beyond the scope of this work.

On the other hand, major components of the organic materials in carbonaceous chondrites are composed of insoluble organic matter (IOM), which is a highly C-rich macromolecule.
As shown in Sect.~\ref{sec:d_tem}, when $T \ga 140$ K, the carbon molecules tend to form C-C bond, resulting in IOM-like molecules.
This bonding change may correspond to the alteration from primitive COMs to IOM, indicating the importance of temperature history for the chemical evolution in protoplanetary disks.

Although not considered in this study, molecular desorption also becomes increasingly important with rising temperatures.
Since the H atom is the smallest mass atom, it may be selectively desorbed.
In that case, the parameter of molecular complexity $\Psi$ (Eq.~(\ref{eq:Psi})) would increases as the desorption proceeds, allowing for the formation of amino acids and sugars even if the initial ice composition is extremely H-rich.
On the other hand, too large $\Psi$ lead to the formation of large molecules depleted of H atoms, with C, N, and O atoms connected to each other (Fig.~\ref{fig:final_products}), thereby reducing molecular diversity.
The moderate value of $\Psi \sim 0.5$-1.0 (except for too large and too small C/O ratio), which maximizes the final abundances of amino acids and sugars (Sect.~\ref{sec:semi_analy}), would also correspond to the optimal condition for the production of diverse COMs.
For COM synthesis, it is imperative to carefully evaluate the compositional changes that occur during UV irradiation.

The experienced temperature of the icy dust grains is thus crucial for the organic material evolution of protoplanetary disks.
On the other hand, due to the weak timescale dependence (Sect.~\ref{sec:d_time}), COMs in ice would be maintained on very long timescales unless decomposed or desorbed.
As the disk evolves, some COMs, including amino acids and sugars, would be incorporated into small bodies.

\section{Conclusions}
\label{sec:conclusion}

In this study, we investigate UV-driven synthesis of COMs, particularly amino acids and sugars, on ice surfaces by means of a novel chemical reaction simulation.
Aiming to explore global reaction networks of COM synthesis, which are not easy to address by experiments or conventional theoretical calculations, we have developed a computational model that does not assume reaction pathways in advance.
In this model, using a Monte Carlo method, chemical evolution within a given molecular set is obtained by repeatedly selecting reactions according to weighted probabilities based on activation energy.

This model was designed to significantly reduce computational costs by adopting very approximate calculations of activation energies, instead of accurate quantum chemical calculations.
This has allowed for the reaction simulation containing a large number of atoms with a broad range of parameters and initial conditions.
In this study, we have also developed explicit prescriptions of radical reactions, CO molecules, and timescales of the reactions, in order to reproduce more realistic chemical reactions.

This study is focused on the COMs' synthesis occurring in icy dust grains in a protoplanetary disk.
Assuming a situation in which the grains wind up in the upper layer are temporarily exposed to ultraviolet radiation and then sink into the disk, two phases were set in the simulations: UV phase and post-UV phase.
The reaction processes and molecular evolution of the initial molecules consisting of methanol, formaldehyde, ammonia, and water molecules through these phases are calculated  at $T =100$ K.
The main results in this study are summarized as follows.
\begin{enumerate}
      \item During the UV phase, 
      photodissociation and radical-radical reactions randomize the covalent bonds between C, N, O, and H atoms within the intial molecules, providing various bonds.
      Due to the randomization, the final abundances of the products, including amino acids and sugars, are essentially determined by the atomic ratios of initial molecules such as C/H and O/H.
      \item Reactions in the post-UV phase are dominated by the barrieless radical-radical reactions at first and gradually shift to the reactions with a low activation barrier. 
      The reactions are finally terminated due to the low temperature and the restriction of reaction timescales imposed by icy grain motions in the disk.
      \item 
      The synthesis of amino acids and sugars, which have been investigated separately in previous studies, were confirmed within the same simulation.
      They both were not formed by build-up-type reactions such as the Strecker reaction and the formose reaction, but through the random bond formation by radical reactions.
      \item The calculations with 57 different sets of initial molecules showed that the final abundance of amino acids is generally more than ten times higher than that of sugars.
      \item Their abundances peaked at $\rm C/H \sim 0.1$-0.3 and $\rm O/H \sim 0.3$-0.5. To understand the abundance distributions on the C/H-O/H plane, a semi-analytical formula was derived, based on the four inhibitory mechanisms for their formation.
      \item We introduced the index ($\Psi$) of potential complexity of a molecular set and found that the synthesis of amino acids and sugars is optimized at $\Psi \sim 1$, while only simple molecules (CH$_4$, H$_2$O and NH$_3$) remain at $\Psi \ll 1$ and IOM-like large complex molecules are formed for $\Psi \ga 2-3$.
      \item 
      The results were not sensitive to UV photon energy, the uncertainty in the Bell-Evans-Polanyi principle, and the upper limit of reaction timescales.
      An important finding is that lower temperatures are favorable to preserve the final yields of amino acids and sugars against decomposition.
      \end{enumerate}

Our current simulation assumes a closed system and a condition of efficient photodissociation and diffusion independent of molecular species.
By eliminating the complicating factors that occur in reality, we have deduced the  important suggestions noted above for the relatively simple and intrinsic mechanisms of the synthesis of amino acids and sugars.  
To compare our simulation with real systems, more comprehensive in situ analysis data from laboratory experiments, observations of icy dust composition, and further improvements to our theoretical model are needed.

\begin{acknowledgements}
We thank Yuichiro Ueno, Yoko Kebukawa, Yasuhiro Oba, Yuri Aikawa, Jim Cleaves, Markus Meringer, Shwan McGlynn, Kosuke Kurosawa, Shogo Tachibana, Yu Komatsu, Hideko Nomura, Yasuhito Sekine and Katsuyuki Kawamura for fruitful discussions and constructive suggestions. This work was supported by JSPS Kakenhi 21H04512.
\end{acknowledgements}

%
%

\bibliography{main}{}

\begin{thebibliography}{77}
\expandafter\ifx\csname natexlab\endcsname\relax\def\natexlab#1{#1}\fi

\bibitem[{{Altwegg} {et~al.}(2016){Altwegg}, {Balsiger}, {Bar-Nun}, {Berthelier}, {Bieler}, {Bochsler}, {Briois}, {Calmonte}, {Combi}, {Cottin}, {De Keyser}, {Dhooghe}, {Fiethe}, {Fuselier}, {Gasc}, {Gombosi}, {Hansen}, {Haessig}, {Ja ckel}, {Kopp}, {Korth}, {Le Roy}, {Mall}, {Marty}, {Mousis}, {Owen}, {Reme}, {Rubin}, {Semon}, {Tzou}, {Waite}, \& {Wurz}}]{Altwegg2016}
{Altwegg}, K., {Balsiger}, H., {Bar-Nun}, A., {et~al.} 2016, Science Advances, 2, e1600285

\bibitem[{Arumainayagam {et~al.}(2019)Arumainayagam, Garrod, Boyer, Hay, Bao, Campbell, Wang, Nowak, Arumainayagam, \& Hodge}]{Arumainayagam2019}
Arumainayagam, C.~R., Garrod, R.~T., Boyer, M.~C., {et~al.} 2019, Chem. Soc. Rev., 48, 2293

\bibitem[{Azofra {et~al.}(2012)Azofra, Alkorta, Elguero, \& Popelier}]{Azofra2012}
Azofra, L.~M., Alkorta, I., Elguero, J., \& Popelier, P.~L. 2012, Carbohydrate Research, 358, 96

\bibitem[{{Bekaert} {et~al.}(2018){Bekaert}, {Derenne}, {Tissandier}, {Marrocchi}, {Charnoz}, {Anquetil}, \& {Marty}}]{Bekaert2018}
{Bekaert}, D.~V., {Derenne}, S., {Tissandier}, L., {et~al.} 2018, \apj, 859, 142

\bibitem[{Bulak {et~al.}(2021)Bulak, Paardekooper, Fedoseev, \& Linnartz}]{Bulak2021}
Bulak, M., Paardekooper, D.~M., Fedoseev, G., \& Linnartz, H. 2021, \aap, 647, A82

\bibitem[{{Ceccarelli} {et~al.}(2023){Ceccarelli}, {Codella}, {Balucani}, {Bockelee-Morvan}, {Herbst}, {Vastel}, {Caselli}, {Favre}, {Lefloch}, {Oberg}, \& {Yamamoto}}]{PP7_ceccarrelli}
{Ceccarelli}, C., {Codella}, C., {Balucani}, N., {et~al.} 2023, in Astronomical Society of the Pacific Conference Series, Vol. 534, Astronomical Society of the Pacific Conference Series, ed. S.~{Inutsuka}, Y.~{Aikawa}, T.~{Muto}, K.~{Tomida}, \& M.~{Tamura}, 379

\bibitem[{Chang \& Herbst(2016)}]{chang_2016}
Chang, Q. \& Herbst, E. 2016, \apj, 819, 145

\bibitem[{{Ciesla} \& {Sandford}(2012)}]{ciesla2012}
{Ciesla}, F.~J. \& {Sandford}, S.~A. 2012, Science, 336, 452

\bibitem[{{Danger} {et~al.}(2011){Danger}, {Borget}, {Chomat}, {Duvernay}, {Theul{\'e}}, {Guillemin}, {Le Sergeant D'Hendecourt}, \& {Chiavassa}}]{Danger2011}
{Danger}, G., {Borget}, F., {Chomat}, M., {et~al.} 2011, \aap, 535, A47

\bibitem[{{Dass} {et~al.}(2021){Dass}, {Georgelin}, {Westall}, {Foucher}, {De Los Rios}, {Busiello}, {Liang}, \& {Piazza}}]{Dass2021}
{Dass}, A.~V., {Georgelin}, T., {Westall}, F., {et~al.} 2021, Nature Communications, 12, 2749

\bibitem[{{de Marcellus} {et~al.}(2015){de Marcellus}, {Meinert}, {Myrgorodska}, {Nahon}, {Buhse}, {d'Hendecourt}, \& {Meierhenrich}}]{Marcellus2015}
{de Marcellus}, P., {Meinert}, C., {Myrgorodska}, I., {et~al.} 2015, Proceedings of the National Academy of Science, 112, 965

\bibitem[{Dugundji \& Ugi(1973)}]{dugundji}
Dugundji, J. \& Ugi, I. 1973, Computers in Chemistry, 19

\bibitem[{Ehrenfreund {et~al.}(2001)Ehrenfreund, Glavin, Botta, Cooper, \& Bada}]{Ehrenfreund2001}
Ehrenfreund, P., Glavin, D.~P., Botta, O., Cooper, G., \& Bada, J.~L. 2001, Proc. Natl. Acad. Sci. U. S. A., 98, 2138

\bibitem[{Enrique-Romero {et~al.}(2022)Enrique-Romero, Rimola, Ceccarelli, Ugliengo, Balucani, \& Skouteris}]{Enrique-Romero2022-mv}
Enrique-Romero, J., Rimola, A., Ceccarelli, C., {et~al.} 2022, Astrophys. J. Suppl. Ser., 259, 39

\bibitem[{{Fresneau} {et~al.}(2015){Fresneau}, {Danger}, {Rimola}, {Duvernay}, {Theul{\'e}}, \& {Chiavassa}}]{Fresneau2015}
{Fresneau}, A., {Danger}, G., {Rimola}, A., {et~al.} 2015, \mnras, 451, 1649

\bibitem[{Furukawa {et~al.}(2019)Furukawa, Chikaraishi, Ohkouchi, Ogawa, Glavin, Dworkin, Abe, \& Nakamura}]{furukawa_2019}
Furukawa, Y., Chikaraishi, Y., Ohkouchi, N., {et~al.} 2019, Proceedings of the National Academy of Sciences, 116, 24440

\bibitem[{{Garrod}(2019)}]{garrod_2019}
{Garrod}, R.~T. 2019, \apj, 884, 69

\bibitem[{{Garrod} \& {Herbst}(2006)}]{garrod_herbst2006}
{Garrod}, R.~T. \& {Herbst}, E. 2006, \aap, 457, 927

\bibitem[{Gibb {et~al.}(2004)Gibb, Whittet, Boogert, \& Tielens}]{Gibb2004-ei}
Gibb, E.~L., Whittet, D. C.~B., Boogert, A. C.~A., \& Tielens, A. G. G.~M. 2004, Astrophys. J. Suppl. Ser., 151, 35

\bibitem[{{Goesmann} {et~al.}(2015){Goesmann}, {Rosenbauer}, {Bredeh{\"o}ft}, {Cabane}, {Ehrenfreund}, {Gautier}, {Giri}, {Kr{\"u}ger}, {Le Roy}, {MacDermott}, {McKenna-Lawlor}, {Meierhenrich}, {Caro}, {Raulin}, {Roll}, {Steele}, {Steininger}, {Sternberg}, {Szopa}, {Thiemann}, \& {Ulamec}}]{Goesmann2015}
{Goesmann}, F., {Rosenbauer}, H., {Bredeh{\"o}ft}, J.~H., {et~al.} 2015, Science, 349, 2.689

\bibitem[{Goldman {et~al.}(2010)Goldman, Reed, Fried, William~Kuo, \& Maiti}]{Goldman2010}
Goldman, N., Reed, E.~J., Fried, L.~E., William~Kuo, I.-F., \& Maiti, A. 2010, Nat. Chem., 2, 949

\bibitem[{{Gudipati} \& {Yang}(2012)}]{Gudipati2012}
{Gudipati}, M.~S. \& {Yang}, R. 2012, \apjl, 756, L24

\bibitem[{Habershon(2015)}]{Habershon_2015}
Habershon, S. 2015, J Chem Phys, 143, 094106

\bibitem[{{H{\"a}nni} {et~al.}(2020){H{\"a}nni}, {Altwegg}, {Pestoni}, {Rubin}, {Schroeder}, {Schuhmann}, \& {Wampfler}}]{Hanni2020}
{H{\"a}nni}, N., {Altwegg}, K., {Pestoni}, B., {et~al.} 2020, \mnras, 498, 2239

\bibitem[{{Heays} {et~al.}(2017){Heays}, {Bosman}, \& {van Dishoeck}}]{Heays2017}
{Heays}, A.~N., {Bosman}, A.~D., \& {van Dishoeck}, E.~F. 2017, \aap, 602, A105

\bibitem[{{Henderson} \& {Gudipati}(2015)}]{Henderson2015}
{Henderson}, B.~L. \& {Gudipati}, M.~S. 2015, \apj, 800, 66

\bibitem[{{Herbst} \& {van Dishoeck}(2009)}]{Herbst2009}
{Herbst}, E. \& {van Dishoeck}, E.~F. 2009, \araa, 47, 427

\bibitem[{Inostroza {et~al.}(2019)Inostroza, Mardones, Cernicharo, Zinnecker, Ge, Aria, Fuentealba, \& Cardenas}]{Inostroza2019}
Inostroza, N., Mardones, D., Cernicharo, J., {et~al.} 2019, \aap, 629, A28

\bibitem[{Jin \& Garrod(2020)}]{jin_2020}
Jin, M. \& Garrod, R.~T. 2020, The Astrophysical Journal Supplement Series, 249, 26

\bibitem[{J{\o}rgensen {et~al.}(2020)J{\o}rgensen, Belloche, \& Garrod}]{Jorgensen2020}
J{\o}rgensen, J.~K., Belloche, A., \& Garrod, R.~T. 2020, Annu. Rev. Astron. Astrophys., 58, 727

\bibitem[{Kebukawa {et~al.}(2022)Kebukawa, Asano, Tani, Yoda, \& Kobayashi}]{Kebukawa2022}
Kebukawa, Y., Asano, S., Tani, A., Yoda, I., \& Kobayashi, K. 2022, ACS Cent. Sci., 8, 1664

\bibitem[{{Klarmann} {et~al.}(2018){Klarmann}, {Ormel}, \& {Dominik}}]{Klarmann2018}
{Klarmann}, L., {Ormel}, C.~W., \& {Dominik}, C. 2018, \aap, 618, L1

\bibitem[{{Koga} \& {Naraoka}(2017)}]{Koga2017}
{Koga}, T. \& {Naraoka}, H. 2017, Scientific Reports, 7, 636

\bibitem[{{Koga} \& {Naraoka}(2022)}]{Koga2022}
{Koga}, T. \& {Naraoka}, H. 2022, ACS Earth and Space Chemistry, 6, 1311

\bibitem[{{Kuga} {et~al.}(2015){Kuga}, {Marty}, {Marrocchi}, \& {Tissandier}}]{Kuga2015}
{Kuga}, M., {Marty}, B., {Marrocchi}, Y., \& {Tissandier}, L. 2015, Proceedings of the National Academy of Science, 112, 7129

\bibitem[{Magrino {et~al.}(2021)Magrino, Pietrucci, \& Saitta}]{Magrino2021}
Magrino, T., Pietrucci, F., \& Saitta, A.~M. 2021, J. Phys. Chem. Lett., 12, 2630

\bibitem[{{Mart{\'\i}n-Dom{\'e}nech} {et~al.}(2020){Mart{\'\i}n-Dom{\'e}nech}, {{\"O}berg}, \& {Rajappan}}]{Martin2020}
{Mart{\'\i}n-Dom{\'e}nech}, R., {{\"O}berg}, K.~I., \& {Rajappan}, M. 2020, \apj, 894, 98

\bibitem[{Mart{\'\i}nez-Bachs \& Rimola(2023)}]{Martinez-Bachs2023-sz}
Mart{\'\i}nez-Bachs, B. \& Rimola, A. 2023, Int. J. Mol. Sci., 24

\bibitem[{{McClure} {et~al.}(2023){McClure}, {Rocha}, {Pontoppidan}, {Crouzet}, {Chu}, {Dartois}, {Lamberts}, {Noble}, {Pendleton}, {Perotti}, {Qasim}, {Rachid}, {Smith}, {Sun}, {Beck}, {Boogert}, {Brown}, {Caselli}, {Charnley}, {Cuppen}, {Dickinson}, {Drozdovskaya}, {Egami}, {Erkal}, {Fraser}, {Garrod}, {Harsono}, {Ioppolo}, {Jim{\'e}nez-Serra}, {Jin}, {J{\o}rgensen}, {Kristensen}, {Lis}, {McCoustra}, {McGuire}, {Melnick}, {{\~A}-berg}, {Palumbo}, {Shimonishi}, {Sturm}, {van Dishoeck}, \& {Linnartz}}]{McClure2023}
{McClure}, M.~K., {Rocha}, W.~R.~M., {Pontoppidan}, K.~M., {et~al.} 2023, Nature Astronomy, 7, 431

\bibitem[{{McKay} \& {Roth}(2021)}]{McKay2021}
{McKay}, A.~J. \& {Roth}, N.~X. 2021, Life, 11, 37

\bibitem[{{Meinert} {et~al.}(2016){Meinert}, {Myrgorodska}, {de Marcellus}, {Buhse}, {Nahon}, {Hoffmann}, {d'Hendecourt}, \& {Meierhenrich}}]{Meinert2016}
{Meinert}, C., {Myrgorodska}, I., {de Marcellus}, P., {et~al.} 2016, Science, 352, 208

\bibitem[{Michaelides {et~al.}(2003)Michaelides, Liu, Zhang, Alavi, King, \& Hu}]{michaelides_2003}
Michaelides, A., Liu, Z.-P., Zhang, C.~J., {et~al.} 2003, Journal of the American Chemical Society, 125, 3704, pMID: 12656593

\bibitem[{{Molpeceres} {et~al.}(2023){Molpeceres}, {Enrique-Romero}, \& {Aikawa}}]{Molpeceres2023}
{Molpeceres}, G., {Enrique-Romero}, J., \& {Aikawa}, Y. 2023, \aap, 677, A39

\bibitem[{{Mu{\~n}oz Caro} {et~al.}(2002){Mu{\~n}oz Caro}, {Meierhenrich}, {Schutte}, {Barbier}, {Arcones Segovia}, {Rosenbauer}, {Thiemann}, {Brack}, \& {Greenberg}}]{Munoz2002}
{Mu{\~n}oz Caro}, G.~M., {Meierhenrich}, U.~J., {Schutte}, W.~A., {et~al.} 2002, \nat, 416, 403

\bibitem[{{Mu{\~n}oz Caro} \& {Schutte}(2003)}]{Munoz2003}
{Mu{\~n}oz Caro}, G.~M. \& {Schutte}, W.~A. 2003, \aap, 412, 121

\bibitem[{Mu{\~n}oz~Caro {et~al.}(2019)Mu{\~n}oz~Caro, Ciaravella, Jim{\'e}nez-Escobar, Cecchi-Pestellini, Gonz{\'a}lez-D{\'\i}az, \& Chen}]{Munoz_Caro2019}
Mu{\~n}oz~Caro, G.~M., Ciaravella, A., Jim{\'e}nez-Escobar, A., {et~al.} 2019, ACS Earth Space Chem., 3, 2138

\bibitem[{Naraoka {et~al.}(2023)Naraoka, Takano, Dworkin, Oba, Hamase, Furusho, Ogawa, Hashiguchi, Fukushima, Aoki, Schmitt-Kopplin, Aponte, Parker, Glavin, McLain, Elsila, Graham, Eiler, Orthous-Daunay, Wolters, Isa, Vuitton, Thissen, Sakai, Yoshimura, Koga, Ohkouchi, Chikaraishi, Sugahara, Mita, Furukawa, Hertkorn, Ruf, Yurimoto, Nakamura, Noguchi, Okazaki, Yabuta, Sakamoto, Tachibana, Connolly, Lauretta, Abe, Yada, Nishimura, Yogata, Nakato, Yoshitake, Suzuki, Miyazaki, Furuya, Hatakeda, Soejima, Hitomi, Kumagai, Usui, Hayashi, Yamamoto, Fukai, Kitazato, Sugita, Namiki, Arakawa, Ikeda, Ishiguro, Hirata, Wada, Ishihara, Noguchi, Morota, Sakatani, Matsumoto, Senshu, Honda, Tatsumi, Yokota, Honda, Michikami, Matsuoka, Miura, Noda, Yamada, Yoshihara, Kawahara, Ozaki, Iijima, Yano, Hayakawa, Iwata, Tsukizaki, Sawada, Hosoda, Ogawa, Okamoto, Hirata, Shirai, Shimaki, Yamada, Okada, Yamamoto, Takeuchi, Fujii, Takei, Yoshikawa, Mimasu, Ono, Ogawa, Kikuchi, Nakazawa, Terui, Tanaka, Saiki, Yoshikawa, Watanabe, \&
  Tsuda}]{Naraoka2023}
Naraoka, H., Takano, Y., Dworkin, J.~P., {et~al.} 2023, Science, 379, eabn9033

\bibitem[{{Nuevo} {et~al.}(2008){Nuevo}, {Auger}, {Blanot}, \& {D'Hendecourt}}]{Nuevo2008}
{Nuevo}, M., {Auger}, G., {Blanot}, D., \& {D'Hendecourt}, L. 2008, Origins of Life and Evolution of the Biosphere, 38, 37

\bibitem[{{Oba} {et~al.}(2019){Oba}, {Takano}, {Naraoka}, {Watanabe}, \& {Kouchi}}]{Oba2019}
{Oba}, Y., {Takano}, Y., {Naraoka}, H., {Watanabe}, N., \& {Kouchi}, A. 2019, Nature Communications, 10, 4413

\bibitem[{{Oba} {et~al.}(2010){Oba}, {Watanabe}, {Kouchi}, {Hama}, \& {Pirronello}}]{Oba2010}
{Oba}, Y., {Watanabe}, N., {Kouchi}, A., {Hama}, T., \& {Pirronello}, V. 2010, \apjl, 712, L174

\bibitem[{{\"O}berg(2016)}]{oberg2016}
{\"O}berg, K.~I. 2016, Chem. Rev., 116, 9631

\bibitem[{{{\"O}berg} {et~al.}(2009){{\"O}berg}, {Garrod}, {van Dishoeck}, \& {Linnartz}}]{Oberg2009}
{{\"O}berg}, K.~I., {Garrod}, R.~T., {van Dishoeck}, E.~F., \& {Linnartz}, H. 2009, \aap, 504, 891

\bibitem[{Pantaleone {et~al.}(2021)Pantaleone, Enrique-Romero, Ceccarelli, Ferrero, Balucani, Rimola, \& Ugliengo}]{Pantaleone2021-tn}
Pantaleone, S., Enrique-Romero, J., Ceccarelli, C., {et~al.} 2021, \apj, 917, 49

\bibitem[{{Parker} {et~al.}(2023){Parker}, {McLain}, {Glavin}, {Dworkin}, {Elsila}, {Aponte}, {Naraoka}, {Takano}, {Tachibana}, {Yabuta}, {Yurimoto}, {Sakamoto}, {Yada}, {Nishimura}, {Nakato}, {Miyazaki}, {Yogata}, {Abe}, {Okada}, {Usui}, {Yoshikawa}, {Saiki}, {Tanaka}, {Nakazawa}, {Tsuda}, {Terui}, {Noguchi}, {Okazaki}, {Watanabe}, \& {Nakamura}}]{Parker2023}
{Parker}, E.~T., {McLain}, H.~L., {Glavin}, D.~P., {et~al.} 2023, \gca, 347, 42

\bibitem[{Pizzarello \& Shock(2010)}]{Pizzarello2010}
Pizzarello, S. \& Shock, E. 2010, Cold Spring Harb. Perspect. Biol., 2, a002105

\bibitem[{Rocha {et~al.}(2024)Rocha, van Dishoeck, Ressler, van Gelder, Slavicinska, Brunken, Linnartz, Ray, Beuther, Caratti~o Garatti, Geers, Kavanagh, Klaassen, Justannont, Chen, Francis, Gieser, Perotti, Tychoniec, Barsony, Majumdar, le~Gouellec, Chu, Lew, Henning, \& Wright}]{Rocha2024}
Rocha, W. R.~M., van Dishoeck, E.~F., Ressler, M.~E., {et~al.} 2024, \aap

\bibitem[{{Saitta} \& {Saija}(2014)}]{saitta_2014}
{Saitta}, A.~M. \& {Saija}, F. 2014, Proceedings of the National Academy of Science, 111, 13768

\bibitem[{Sameera {et~al.}(2022)Sameera, Senevirathne, Nguyen, Oba, Ishibashi, Tsuge, Hidaka, \& Watanabe}]{Sameera2022}
Sameera, W. M.~C., Senevirathne, B., Nguyen, T., {et~al.} 2022, Front. Astron. Space Sci., 9

\bibitem[{Sanderson(1976)}]{sanderson}
Sanderson, R. 1976, Chemical bonds and bonds energy, Vol.~21 (Elsevier)

\bibitem[{Schr{\"o}der {et~al.}(1998)Schr{\"o}der, Heinemann, Schwarz, Harvey, Dua, Blanksby, \& Bowie}]{Schroder1998}
Schr{\"o}der, D., Heinemann, C., Schwarz, H., {et~al.} 1998, Chemistry--A European Journal, 4, 2550

\bibitem[{Sephton(2002)}]{Sephton2002}
Sephton, M.~A. 2002, Nat. Prod. Rep., 19, 292

\bibitem[{Sewi{\l}o {et~al.}(2022)Sewi{\l}o, Cordiner, Charnley, Oliveira, Garcia-Berrios, Schilke, Ward, Wiseman, Indebetouw, Tokuda, van Loon, S{\'a}nchez-Monge, Allen, Chen, Hamedani~Golshan, Karska, Kristensen, Kurtz, M{\"o}ller, Onishi, \& Zahorecz}]{Sewilo2022-ey}
Sewi{\l}o, M., Cordiner, M., Charnley, S.~B., {et~al.} 2022, \apj, 931, 102

\bibitem[{{Singh} {et~al.}(2022){Singh}, {Zhu}, {La Jeunesse}, {Fortenberry}, \& {Kaiser}}]{Singh2022}
{Singh}, S.~K., {Zhu}, C., {La Jeunesse}, J., {Fortenberry}, R.~C., \& {Kaiser}, R.~I. 2022, Nature Communications, 13, 375

\bibitem[{{Sugahara} {et~al.}(2019){Sugahara}, {Takano}, {Tachibana}, {Sugawara}, {Chikaraishi}, {Ogawa}, {Ohkouchi}, {Kouchi}, \& {Yurimoto}}]{Sugahara2019}
{Sugahara}, H., {Takano}, Y., {Tachibana}, S., {et~al.} 2019, Geochemical Journal, 53, 5

\bibitem[{Sutton \& Vlachos(2012)}]{sutton_2012}
Sutton, J.~E. \& Vlachos, D.~G. 2012, ACS Catalysis, 2, 1624

\bibitem[{{Tachibana} {et~al.}(2017){Tachibana}, {Kouchi}, {Hama}, {Oba}, {Piani}, {Sugawara}, {Endo}, {Hidaka}, {Kimura}, {Murata}, {Yurimoto}, \& {Watanabe}}]{tachibana}
{Tachibana}, S., {Kouchi}, A., {Hama}, T., {et~al.} 2017, Science Advances, 3, eaao2538

\bibitem[{{Takehara} {et~al.}(2022){Takehara}, {Shoji}, \& {Ida}}]{takehara}
{Takehara}, H., {Shoji}, D., \& {Ida}, S. 2022, \aap, 662, A76

\bibitem[{Tsuge {et~al.}(2023)Tsuge, Molpeceres, Aikawa, \& Watanabe}]{Tsuge2023}
Tsuge, M., Molpeceres, G., Aikawa, Y., \& Watanabe, N. 2023, Nature Astronomy, 7, 1351

\bibitem[{Turner {et~al.}(2021)Turner, Chandra, Fortenberry, \& Kaiser}]{Turner2021}
Turner, A.~M., Chandra, S., Fortenberry, R.~C., \& Kaiser, R.~I. 2021, Chemphyschem, 22, 985

\bibitem[{Turner \& Kaiser(2020)}]{Turner2020}
Turner, A.~M. \& Kaiser, R.~I. 2020, Acc. Chem. Res., 53, 2791

\bibitem[{van Santen {et~al.}(2010)van Santen, Neurock, \& Shetty}]{Santen2010}
van Santen, R.~A., Neurock, M., \& Shetty, S.~G. 2010, Chem. Rev., 110, 2005

\bibitem[{Wang {et~al.}(2011)Wang, Temel, Shen, Jones, Grabow, Studt, Bligaard, Abild-Pedersen, Christensen, \& N{¥o}rskov}]{Wang_2011_evans}
Wang, S., Temel, B., Shen, J., {et~al.} 2011, Catalysis Letters, 141, 370

\bibitem[{Yabuta {et~al.}(2023)Yabuta, Cody, Engrand, Kebukawa, De~Gregorio, Bonal, Remusat, Stroud, Quirico, Nittler, Hashiguchi, Komatsu, Okumura, Mathurin, Dartois, Duprat, Takahashi, Takeichi, Kilcoyne, Yamashita, Dazzi, Deniset-Besseau, Sandford, Martins, Tamenori, Ohigashi, Suga, Wakabayashi, Verdier-Paoletti, Mostefaoui, Montagnac, Barosch, Kamide, Shigenaka, Bejach, Matsumoto, Enokido, Noguchi, Yurimoto, Nakamura, Okazaki, Naraoka, Sakamoto, Connolly, Lauretta, Abe, Okada, Yada, Nishimura, Yogata, Nakato, Yoshitake, Iwamae, Furuya, Hatakeda, Miyazaki, Soejima, Hitomi, Kumagai, Usui, Hayashi, Yamamoto, Fukai, Sugita, Kitazato, Hirata, Honda, Morota, Tatsumi, Sakatani, Namiki, Matsumoto, Noguchi, Wada, Senshu, Ogawa, Yokota, Ishihara, Shimaki, Yamada, Honda, Michikami, Matsuoka, Hirata, Arakawa, Okamoto, Ishiguro, Jaumann, Bibring, Grott, Schr{\"o}der, Otto, Pilorget, Schmitz, Biele, Ho, Moussi-Soffys, Miura, Noda, Yamada, Yoshihara, Kawahara, Ikeda, Yamamoto, Shirai, Kikuchi, Ogawa, Takeuchi, Ono,
  Mimasu, Yoshikawa, Takei, Fujii, Iijima, Nakazawa, Hosoda, Iwata, Hayakawa, Sawada, Yano, Tsukizaki, Ozaki, Terui, Tanaka, Fujimoto, Yoshikawa, Saiki, Tachibana, Watanabe, \& Tsuda}]{Yabuta2023}
Yabuta, H., Cody, G.~D., Engrand, C., {et~al.} 2023, Science, 379, eabn9057

\bibitem[{Yamato {et~al.}(2024)Yamato, Notsu, Aikawa, Okoda, Nomura, \& Sakai}]{Yamato2024-hs}
Yamato, Y., Notsu, S., Aikawa, Y., {et~al.} 2024, Astron. J., 167, 66

\bibitem[{{Yang} {et~al.}(2022){Yang}, {Green}, {Pontoppidan}, {Bergner}, {Cleeves}, {Evans}, {Garrod}, {Jin}, {Kim}, {Kim}, {Lee}, {Sakai}, {Shingledecker}, {Shope}, {Tobin}, \& {van Dishoeck}}]{Yang2022}
{Yang}, Y.-L., {Green}, J.~D., {Pontoppidan}, K.~M., {et~al.} 2022, \apjl, 941, L13

\bibitem[{Zamirri {et~al.}(2019)Zamirri, Ugliengo, Ceccarelli, \& Rimola}]{Zamirri2019}
Zamirri, L., Ugliengo, P., Ceccarelli, C., \& Rimola, A. 2019, ACS Earth Space Chem., 3, 1499

\bibitem[{Zhu {et~al.}(2021)Zhu, Bergantini, Singh, Abplanalp, \& Kaiser}]{Zhu2021}
Zhu, C., Bergantini, A., Singh, S.~K., Abplanalp, M.~J., \& Kaiser, R.~I. 2021, \apj, 920, 73

\end{thebibliography}

\appendix

\section{Initial molecular sets used for Fig.~\ref{fig:contour}.}

\begin{table}[h]
    \centering
    \fontsize{9pt}{9pt}\selectfont
    \begin{tabular}{c|l|ccc} \hline
    No. & Initial molecular set & $\Psi$ & C/H & O/H \\ \hline\hline
    1 & 2 $\rm CH_3OH$, 5 $\rm CH_2O$, 9 $\rm NH_3$, 22 $\rm H_2O$& 0.27 & 0.079 & 0.33 \\
    2 & 5 $\rm CH_4$, 12 $\rm NH_3$, 25 $\rm H_2O$, 10 $\rm H_2$& 0 & 0.040 & 0.20 \\
    3 & 9 $\rm CH_4$, 13 $\rm NH_3$, 27 $\rm H_2O$ & 0 & 0.070 & 0.21 \\
    4 & 3 $\rm CH_4$, 2 $\rm CH_3OH$, 12 $\rm NH_3$, 32 $\rm H_2O$& 0.033 & 0.042 & 0.28 \\
    5 & 10 $\rm CH_3OH$, 11 $\rm NH_3$, 16 $\rm H_2O$& 0.19 & 0.095 & 0.25 \\
    6 & 3 $\rm CH_4$, 10 $\rm CH_3OH$, 10 $\rm NH_3$, 9 $\rm H_2O$& 0.20 & 0.13 & 0.19 \\
    7 & 4 $\rm HCN$, 6 $\rm NH_3$, 37 $\rm H_2O$, $\rm O_2$& 0.30 & 0.042 & 0.41 \\
    8 & 6 $\rm CH_3OH$, 5 $\rm CH_2O$, 10 $\rm NH_3$, 18 $\rm H_2O$& 0.32 & 0.11 & 0.29 \\
    9 & $\rm CH_4$, $\rm CH_3OH$, 9 $\rm NH_3$, 28 $\rm H_2O$, 7 $\rm O_2$& 0.33 & 0.022 & 0.47 \\
    10 & $\rm CH_3OH$, 5 $\rm HCOOH$, 8 $\rm NH_3$, 24 $\rm H_2O$& 0.37 & 0.070 & 0.41 \\
    11 & 9 $\rm CH_3OH$, 4 $\rm CH_2O$, 9 $\rm NH_3$, 9 $\rm H_2O$& 0.38 & 0.15 & 0.25 \\
    12 & 7 $\rm CH_3OH$, 4 $\rm CH_2O$, $\rm HCOOH$, 9 $\rm NH_3$, 13 $\rm H_2O$& 0.40 & 0.13 & 0.29 \\
    13 & 3 $\rm CH_2O$, 4 $\rm HCOOH$, 8 $\rm NH_3$, 23 $\rm H_2O$& 0.43 & 0.083 & 0.40 \\
    14 & 13 $\rm CH_3OH$, 3 $\rm CH_2O$, 9 $\rm NH_3$, $\rm H_2O$& 0.44 & 0.18 & 0.20 \\
    15 & 4 $\rm CH_4$, 4 $\rm CH_3OH$, 5 $\rm HCOOH$, 8 $\rm NH_3$, 10 $\rm H_2O$& 0.44 & 0.15 & 0.28 \\
    16 & 17 $\rm CH_4$, 8 $\rm HCN$, 2 $\rm NH_3$, 9 $\rm H_2O$& 0.48 & 0.25 & 0.090 \\
    17 & 5 $\rm CH_2O$, 4 $\rm HCN$, 4 $\rm NH_3$, 27 $\rm H_2O$& 0.55 & 0.11 & 0.40 \\
    18 & 7 $\rm HCOOH$, 7 $\rm NH_3$, 19 $\rm H_2O$& 0.58 & 0.096 & 0.45 \\
    19 & 6 $\rm CO_2$, 8 $\rm NH_3$, 29 $\rm H_2O$& 0.59 & 0.073 & 0.50 \\
    20 & 5 $\rm CH_3OH$, 9 $\rm CH_2O$, 7 $\rm NH_3$, 7 $\rm H_2O$& 0.63 & 0.19 & 0.29 \\
    21 & 8 $\rm CH_2O$, 3 $\rm HCN$, 5 $\rm NH_3$, 22 $\rm H_2O$& 0.64 & 0.14 & 0.38 \\
    22 & 4 $\rm CH_4$, 3 $\rm HCOOH$, 5 $\rm HCN$, 2 $\rm NH_3$, 20 $\rm H_2O$& 0.66 & 0.16 & 0.36 \\
    23 & 5 $\rm HCOOH$, 7 $\rm NH_3$, 22 $\rm H_2O$, 5 $\rm O_2$& 0.67 & 0.067 & 0.56 \\
    24 & 3 $\rm HCOOH$, 7 $\rm NH_3$, 23 $\rm H_2O$, 8 $\rm O_2$& 0.68 & 0.041 & 0.62 \\
    25 & $\rm CH_2O$, 3 $\rm HCOOH$, 5 $\rm HCN$, 2 $\rm NH_3$, 24 $\rm H_2O$& 0.78 & 0.13 & 0.46 \\
    26 & 4 $\rm CH_3OH$, 11 $\rm CH_2O$, 6 $\rm NH_3$, 4 $\rm H_2O$& 0.81 & 0.23 & 0.30 \\
    27 & 4 $\rm CH_2O$, 6 $\rm HCN$, 24 $\rm H_2O$& 0.84 & 0.16 & 0.45 \\
    28 & 11 $\rm CH_2O$, 2 $\rm HCN$, 4 $\rm NH_3$, 14 $\rm H_2O$& 0.88 & 0.20 & 0.39 \\
    29 & 10 $\rm CH_4$, 3 $\rm HCOOH$, 7 $\rm HCN$, 7 $\rm H_2O$& 0.90 & 0.30 & 0.19 \\
    30 & 4 $\rm HCOOH$, $\rm HCN$, 5 $\rm NH_3$, 20 $\rm H_2O$, 7 $\rm O_2$& 0.90 & 0.078 & 0.66 \\
    31 & 6 $\rm HCN$, 27 $\rm H_2O$, 5 $\rm O_2$& 0.93 & 0.10 & 0.62 \\
    32 & 9 $\rm HCOOH$, 6 $\rm NH_3$, 13 $\rm H_2O$, 2 $\rm O_2$& 1 & 0.15 & 0.56 \\
    33 & 2 $\rm CH_2O$, 3 $\rm HCOOH$, 6 $\rm HCN$, 21 $\rm H_2O$& 1.1 & 0.19 & 0.50 \\
    34 & 7 $\rm HCOOH$, 5 $\rm NH_3$, 11 $\rm H_2O$, 5 $\rm O_2$& 1.2 & 0.14 & 0.69 \\
    35 & 10 $\rm CH_4$, 10 $\rm CH_2O$, 7 $\rm HCN$& 1.2 & 0.40 & 0.15 \\
    36 & 9 $\rm HCOOH$, 5 $\rm NH_3$, 9 $\rm H_2O$, 3 $\rm O_2$& 1.3 & 0.18 & 0.65 \\
    37 & $\rm CH_2O$, 10 $\rm HCOOH$, 5 $\rm NH_3$, 6 $\rm H_2O$& 1.3 & 0.22 & 0.55 \\
    38 & 5 $\rm HCN$, 22 $\rm H_2O$, 9 $\rm O_2$& 1.3 & 0.10 & 0.82 \\
    39 & 4 $\rm CH_4$, 8 $\rm C_2H_4$, 6 $\rm CH_2O$, 7 $\rm HCN$& 1.5 & 0.49 & 0.090 \\
    40 & 4 $\rm CH_3OH$, 6 $\rm HCOOH$, 5 $\rm HCN$, 8 $\rm H_2O$& 1.5 & 0.31 & 0.49 \\
    41 & 7 $\rm HCOOH$, 5 $\rm HCN$, 16 $\rm H_2O$, 3 $\rm O_2$& 1.6 & 0.24 & 0.71 \\
    42 & 3 $\rm CH_3OH$, 15 $\rm CH_2O$, 6 $\rm HCN$, 5 $\rm H_2O$& 1.8 & 0.41 & 0.40 \\
    43 & 2 $\rm CH_2O$, 3 $\rm HCOOH$, 5 $\rm HCN$, 9 $\rm C_2H_4$& 1.8 & 0.55 & 0.16 \\
    44 & 3 $\rm CH_2O$, 5 $\rm HCOOH$, 4 $\rm NH_3$, 6 $\rm H_2O$, 8 $\rm O_2$& 1.9 & 0.20 & 0.88 \\
    45 & 10 $\rm CH_2O$, 4 $\rm HCOOH$, 4 $\rm HCN$, 5 $\rm H_2O$& 2.1 & 0.43 & 0.55 \\
    46 & $\rm CH_4$, 18 $\rm CH_2O$, 5 $\rm HCN$, $\rm C_2H_4$& 2.2 & 0.53 & 0.37 \\
    47 & 8 $\rm CO_2$, 4 $\rm HCN$, 18 $\rm H_2O$& 2.2 & 0.30 & 0.85 \\
    48 & 6 $\rm HCOOH$, 4 $\rm CO_2$, 4 $\rm HCN$, 12 $\rm H_2O$& 2.3 & 0.35 & 0.80 \\
    49 & 3 $\rm CH_2O$, 10 $\rm HCOOH$, 4 $\rm HCN$, 5 $\rm H_2O$& 2.4 & 0.43 & 0.70 \\
    50 & 10 $\rm C_3H_4$, 2 $\rm HCOOH$, 5 $\rm HCN$& 2.5 & 0.76 & 0.082 \\
    51 & 13 $\rm CH_2O$, 5 $\rm HCOOH$, 4 $\rm HCN$& 2.7 & 0.55 & 0.58 \\
    52 & 6 $\rm C_3H_4$, 6 $\rm HCOOH$, 4 $\rm HCN$& 2.7 & 0.70 & 0.30 \\
    53 & 5 $\rm CH_2O$, 12 $\rm HCOOH$, 4 $\rm HCN$& 3.1 & 0.55 & 0.76 \\
    54 & 5 $\rm C_3H_4$, 3 $\rm CH_2O$, 5 $\rm HCOOH$, 3 $\rm CO_2$, 4 $\rm HCN$& 3.3 & 0.75 & 0.48 \\
    55 & 5 $\rm C_2H_4$, 9 $\rm CO_2$, 3 $\rm HCN$, 4 $\rm H_2O$& 3.5 & 0.71 & 0.71 \\
    56 & 7 $\rm C_2H_4$, 10 $\rm CO_2$, 3 $\rm HCN$& 4.1 & 0.87 & 0.65 \\
    57 & 13 $\rm CH_2O$, 7 $\rm CO_2$, 3 $\rm HCN$& 4.3 & 0.79 & 0.93 \\ \hline
    \end{tabular}
    
    \caption{The initial molecular sets used to calculate the final abundances of amino acids and sugars in Fig.~\ref{fig:contour}, along with their complexity parameter $\Psi$ (Eq.~(\ref{eq:Psi})), C/H ratio, and O/H ratio. Methane ($\rm CH_4$), ethylene ($\rm C_2H_4$), propadiene ($\rm C_3H_4$), methatnol ($\rm CH_3OH$), formaldelyde ($\rm CH_2O$), formic acid ($\rm HCOOH$), carbon monoxide ($\rm CO_2$), hydrogen cyanide ($\rm HCN$), ammonia ($\rm NH_3$), water ($\rm H_2O$), hydrogen ($\rm H_2$) oxygen ($\rm O_2$) molecules are used as the initial species.
    The N/H ratio of each set is fixed at about 0.1 (0.093-0.105).
    }
    \label{tab:initialsets}
\end{table}

\end{document}